\documentclass[aps,prd,twocolumn,groupedaddress]{revtex4-2}

\usepackage{graphicx,graphics}
\usepackage{dcolumn}
\usepackage{bm}
\usepackage{xcolor}
\usepackage{soul}
\usepackage{hyperref}
\usepackage{amsmath}

\begin{document}


\title{Revisiting pseudo-Dirac neutrino scenario after recent solar neutrino data}


\author{Saeed Ansarifard}
\email{ansarifard@ipm.ir}

\author{Yasaman Farzan}
\email{yasaman@theory.ipm.ac.ir}
\affiliation{School of physics, Institute for Research in Fundamental Sciences (IPM)
\\
P.O.Box 19395-5531, Tehran, Iran}


\date{\today}

\begin{abstract}
It is still unknown whether the mass terms for neutrinos are of Majorana type or of Dirac type. An interesting possibility, known as pseudo-Dirac scheme combines these two with a dominant Dirac mass term and a subdominant Majorana one. As a result, the mass eigenstates come in pairs with a maximal mixing and a small splitting determined by the Majorana mass. This will affect the neutrino oscillation pattern for long baselines. We revisit this scenario employing recent solar neutrino data, including the seasonal variation of the $^7$Be flux recently reported by BOREXINO. We constrain the splitting  using these data and  find  that both the time integrated solar neutrino data and the seasonal variation independently point towards a new pseudo-Dirac solution with nonzero splitting for  $\nu_2$ of  $\Delta m_2^2\simeq 1.5\times 10^{-11}$ eV$^2$. We propose alternative methods to test this new solution. In particular, we point out the importance of measuring the solar neutrino flux at the intermediate energies $1.5~{\rm MeV}<E_\nu<3.5~{\rm MeV}$ (below the Super-Kamiokande detection threshold) as well as a more  precise measurement of the $pep$ flux. The code is available on \href{https://github.com/SaeedAnsarifard/SolarNeutrinos-pseudoDirac.git}{Github} 
\end{abstract}


\def\d{{\rm d}}
\def\Epos{E_{\rm pos}}
\def\ap{\approx}
\def\eff{{\rm eft}}
\def\L{{\cal L}}
\newcommand{\vev}[1]{\langle {#1}\rangle}
\newcommand{\CL}   {C.L.}
\newcommand{\dof}  {d.o.f.}
\newcommand{\eVq}  {\text{EA}^2}
\newcommand{\Sol}  {\textsc{sol}}
\newcommand{\SlKm} {\textsc{sol+kam}}
\newcommand{\Atm}  {\textsc{atm}}
\newcommand{\Chooz}{\textsc{chooz}}
\newcommand{\Dms}  {\Delta m^2_\Sol}
\newcommand{\Dma}  {\Delta m^2_\Atm}
\newcommand{\Dcq}  {\Delta\chi^2}
\newcommand{\nbb}{$\beta\beta_{0\nu}$ }
\newcommand {\be}{\begin{equation}}
\newcommand {\ee}{\end{equation}}
\newcommand {\ba}{\begin{eqnarray}}
\newcommand {\ea}{\end{eqnarray}}
\def\VEV#1{\left\langle #1\right\rangle}
\let\vev\VEV
\def\e6{E(6)}
\def\10{SO(10)}
\def\21{SA(2) $\otimes$ U(1) }
\def\321{$\mathrm{SU(3) \otimes SU(2) \otimes U(1)}$ }
\def\lr{SA(2)$_L \otimes$ SA(2)$_R \otimes$ U(1)}
\def\422{SA(4) $\otimes$ SA(2) $\otimes$ SA(2)}

\def\roughly#1{\mathrel{\raise.3ex\hbox{$#1$\kern-.75em
      \lower1ex\hbox{$\sim$}}}} \def\lsim{\roughly<}
\def\gsim{\roughly>}
\def\ltap{\raisebox{-.4ex}{\rlap{$\sim$}} \raisebox{.4ex}{$<$}}
\def\gtap{\raisebox{-.4ex}{\rlap{$\sim$}} \raisebox{.4ex}{$>$}}
\def\lsim{\raise0.3ex\hbox{$\;<$\kern-0.75em\raise-1.1ex\hbox{$\sim\;$}}}
\def\gsim{\raise0.3ex\hbox{$\;>$\kern-0.75em\raise-1.1ex\hbox{$\sim\;$}}}

\maketitle

\section{Introduction}
Lepton flavor violation is the cornerstone of the modern neutrino physics, having been observed in various neutrino experiments such as solar, atmospheric, reactor and long baseline neutrino experiments.
The three neutrino mass and mixing scheme has been established as the standard  solution to the observed lepton flavor violation in  evolution of neutrino states. It is not however known whether the neutrino mass term also violates lepton number or not. In other words, we do not know if the mass terms for neutrinos are of Majorana type or of Dirac type. 
In general, we can  simultaneously write Majorana ($\mu$) and Dirac mass ($m$) terms for neutrinos. At the limit where Majorana term is much smaller than the  Dirac term ({\it i.e.,} in the limit $\mu \ll m$), the scheme is called  pseudo-Dirac. This limit is of interest from both model building and phenomenological point of view. It is straightforward to show that the mass eigenstates composing the active states will split to  pairs of Majorana states with  maximal mixing and tiny mass squared differences given by $\Delta m^2=\mu m$. For baselines much smaller than $E_\nu/(2\mu m)$, the neutrino oscillation pattern will be similar to what expected for the standard three neutrino mass and mixing scheme. For baselines comparable to $E_\nu/(2\mu m)$ or larger, the active to active  neutrino oscillation probability will be smaller than that expected within the standard scheme as a part of the active flux can oscillate to sterile neutrinos. There is already rich literature on the potential of various neutrino observations to test this scenario. Upcoming terrestrial experiments such as DUNE and JUNO can test  $\Delta m^2 \sim 10^{-5}$ eV$^2$ \cite{Anamiati:2019maf}. The galactic supernova neutrinos can probe $\Delta m^2$ down to $10^{-20}$ eV$^2$ 
\cite{Martinez-Soler:2021unz}. Ultrahigh energy cosmic neutrinos can be sensitive to $\Delta m^2>{\rm few}\times 10^{-18}$ eV$^2$ \cite{Esmaili:2012ac,Joshipura:2013yba,Crocker:1999yw,Esmaili:2009fk,Keranen:2003xd,Crocker:2001zs}. Finally, the solar neutrinos can be sensitive to $\Delta m^2\gsim 10^{-13}$ eV$^2$ \cite{Anamiati:2017rxw}.

The possible effects of  pseudo-Dirac neutrino scheme on solar neutrinos has been already discussed in the literature \cite{Anamiati:2017rxw,deGouvea:2009fp}.  Ref \cite{Anamiati:2017rxw} constrains the splittings of $\nu_1$ and $\nu_2$ and finds a solution at $\sim 10^{-11}$ eV$^2$ for the neutrino data. Since  the publication of Ref.~\cite{Anamiati:2017rxw}, BOREXINO has released more data, with a relatively precise measurement of the $pep$ flux as well as the measurement of seasonal  flux variation.  
Moreover, the Super-Kamiokande data has been updated.
We revisit the pseudo-Dirac scheme with the latest available BOREXINO and Super-Kamiokande solar data, also taking  into account the precise measurement of $\Delta m_{21}^2$ by KamLAND. Similarly to Ref.~\cite{Anamiati:2017rxw}, we  find a solution with  nonzero pseudo-Dirac splitting. We discuss the importance of the precise measurement of $^8$B flux at energies between 1.5~MeV to 3~MeV (that is below the detection threshold of Super-Kamiokande and above the $pep$ line) to test this solution.

In the range that the oscillation length  due to the pseudo-Dirac  mass splitting, $\Delta m^2$, is comparable to the variation of the Earth-Sun distance during a  year (resulting from the eccentricity of the Earth orbit), we expect a signature in the seasonal variation. We examine the recently reported seasonal variation of the $^7$Be flux to search for such a variation. Independently of the time integrated analysis, this data also points towards a pseudo-Dirac solution with the same range of $\Delta m^2$.  We propose a few alternative methods to test this new non-trivial solution.

The paper is organized as follows: In sect.~\ref{osc}, we review the oscillation of pseudo-Dirac neutrinos. This discussion is complemented in the appendix with a focus on matter effects as well as on the eccentricity of the Earth orbit. In sect.~\ref{analysis}, we summarize the  basis of our analysis and define the relevant $\chi^2$ tests. In sect.~\ref{results}, we show the implications of the solar neutrino data for the pseudo-Dirac scheme. The concluding remarks and suggestions for further study are given in sect.~\ref{Dis}.

\section{Oscillation of pseudo-Dirac solar neutrinos\label{osc}}
Within the pseudo-Dirac scheme, the neutrino states $\Psi^T=(\nu_L \ \nu_R)$ have both Dirac mass ($m$) and Majorana mass ($\mu$) terms of form
\be \bar{\Psi} m \Psi+\bar{\Psi}^c \mu \Psi \ \ \ \ \ {\rm with} \ \mu\ll m \label{Mmu}\ee
where $\Psi^c=-i\gamma^2\Psi^*$. In general, both $m$ and $\mu$ are $3\times 3$ matrices in the flavor space. For simplicity, we assume that  $m$ and $\mu$ can be simultaneously diagonalized. Then, as shown in the appendix, each Dirac mass eigenstate, $\nu_i$ splits to two Majorana states with a maximal mixing and a splitting of $\Delta m_i^2=2\mu_i m_i$. Thus, the $\nu_e$ survival probability  and the probability of the conversion of $\nu_e$ into sterile neutrinos can be written as 
\begin{align}\label{PeeBULK}
P_{ee}(E_\nu,L,r) \equiv P(\nu_e\to \nu_e) =   \\ \nonumber
\cos^4\theta_{13} \Biggl(  \cos^2\theta_{12} \cos^2\theta_M & \cos^2(\frac{\Delta m^2_1}{4E_\nu}L)  \\ \nonumber
+ \sin^2\theta_{12} \sin^2\theta_M & \cos^2(\frac{\Delta m^2_2}{4E_\nu}L) \Biggl) \\ \nonumber
+ \sin^4\theta_{13}  \cos^2(\frac{\Delta m^2_3}{4E_\nu}L)
\end{align}
and 
\begin{align}\label{PesBULK}
P_{es}(E_\nu,L,r)  \equiv   \sum_{i} P(\nu_e\to \nu_{s_i})  & =  \\ \nonumber
 \cos^2 \theta_{13} \Biggl( \cos^2\theta_M & \sin^2(\frac{\Delta m^2_1}{4E_\nu}L)  \\ \nonumber
 + \sin^2\theta_M  & \sin^2(\frac{\Delta m^2_2}{4E_\nu}L) \Biggl) \\ \nonumber
  + \sin^2 \theta_{13}  \sin^2(\frac{\Delta m^2_3}{4E_\nu}L) 
\end{align}
where $\theta_M$ is the effective  mixing at the production point of the $\nu_e$ inside the sun (at a distance of $r$ from the Sun center) given by Eq. (\ref{thetM}) in the appendix. $L$ is the distance between Sun and Earth and $E_\nu$ is the neutrino energy.

  \begin{figure}[ht]
  \centering
  	\hspace{0cm}
  	\includegraphics[width=0.5\textwidth, height=0.35\textwidth]{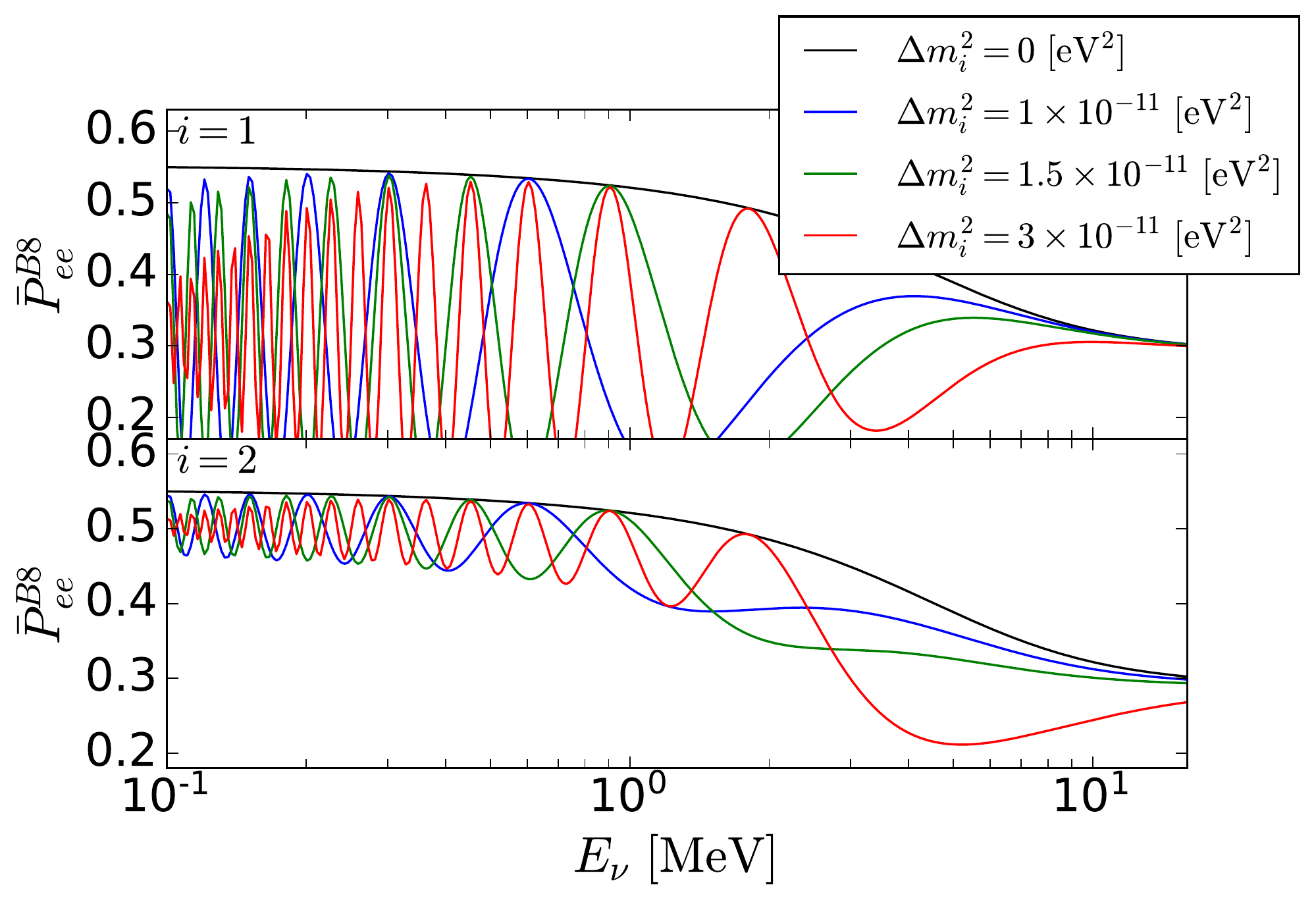}
  	\caption[...]{ Average survival probability of solar neutrinos, $P(\nu_e\to \nu_e$) versus neutrino energy for various values of the splitting.
  		Averaging is taken over the production point assuming the distribution of $^8$B production. {The $L$ dependency is marginalized by taking temporal averaging on a year.} (For more clarification see the bulk of the paper and the appendix.)
  		The black curve corresponds to the pure standard MSW effect. The standard neutrino parameters are set equal to the best fit values as in \cite{Esteban:2020cvm}\footnote{NuFIT 5.1 (2021), \url{www.nu-fit.org}}.  In the upper (lower) panel, we have set $\Delta m_2^2=0$ ($\Delta m_1^2=0$) and only one $\Delta m_i^2$ is set nonzero. 
  		 {The blue, green and red lines  respectively correspond to  $\Delta m_i^2=1\times 10^{-11}$~eV$^2$,  $\Delta m_i^2=1.5\times 10^{-11}$~eV$^2$ and $\Delta m_i^2=3\times 10^{-11}$~eV$^2$.}} 
  	\label{pro}
  \end{figure}  

The density profile of the Sun is exponentially suppressed with the distance from the Sun center so $P_{ee}$ strongly depends on the production point, $r$. As explained in the appendix for each component of the solar flux component ({\it i.e.,} $j \in \{  pp, \ ^7{\rm Be}, \ pep, \ ^8 B \}$) we should average $P_{ee}(E_\nu,L,r)$ and $P_{es}(E_\nu,L,r)$  over the production point, using the production point spatial distribution inside the Sun associated with each flux component. More details can be found in the appendix. Hereafter, we show the averaged survival probability over the production point of component $j$ by $\bar{P}^j_{ee}(E_\nu,L)$. Fig.~\ref{pro} illustrates the oscillation probability averaged with the distribution of $^8$B production at several values of $\Delta m_1^2$ and  $\Delta m_2^2$. The black lines correspond to the standard MSW solution. We expect for relatively large $\Delta m_i^2$, the deviation to be more significant for large energies and relatively suppressed at low energies because the effect is given by the ratio $\Delta m_i^2/E_\nu$.
This behavior is demonstrated by the red lines which correspond to $\Delta m_1^2$ or $\Delta m_2^2=3 \times 10^{-11}$ eV$^2$.  For  $\Delta m_i^2 \sim 10^{-11}$ eV$^2$, the deviation is specially significant for intermediate values of energies $1.5~{\rm MeV}<E_\nu<3~{\rm MeV}$, lying below the detection threshold of Super-Kamiokande where the solar neutrino data is lacking.

  \begin{figure}[ht]
  \centering
  	\hspace{0cm}
  	\includegraphics[width=0.45\textwidth]{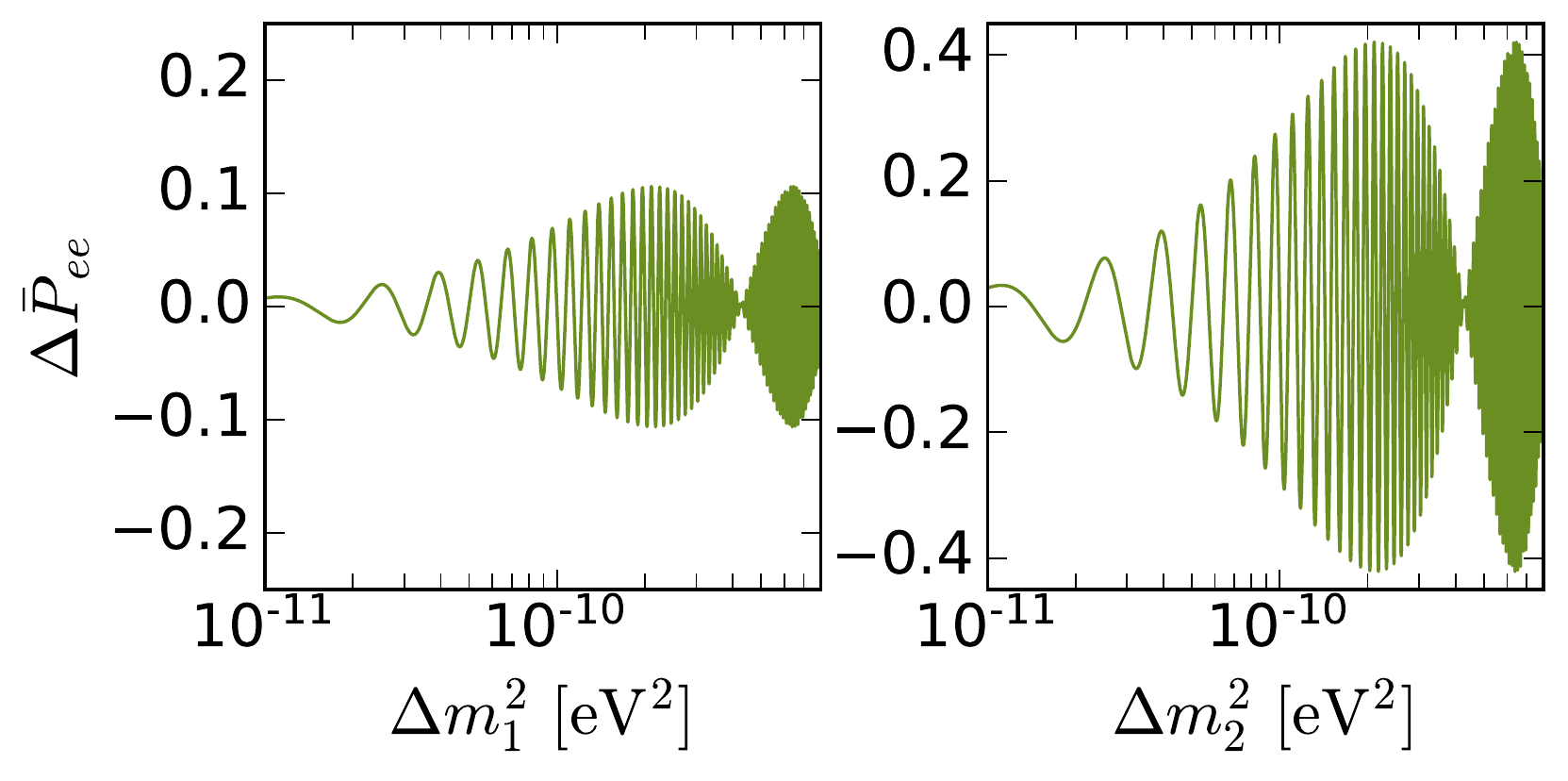}
  	\caption[...]{$\Delta  \bar{P}_{ee} = \bar{P}^{Be7}_{ee}(E_\nu,L_{max})-  \bar{P}^{Be7}_{ee}(E_\nu,L_{min})$ versus $\Delta m_1^2$ and $\Delta m_2^2$ at $E_\nu=0.862$ MeV ({\it i.e.,} for $^7$Be line).  $L_{max}=1.52\times 10^8$~km and  $L_{min}=1.47\times 10^8$~km are the Earth-Sun distance at aphelion and perihelion, respectively.}
  	\label{shaky}
  \end{figure}
  
The average Sun Earth distance is $\sim$ 150 million km but due to the eccentricity of the Earth orbit around the Sun, $L$ varies during a year with $L_{max}-L_{min}=5$ million km. Fig. \ref{shaky} shows $$\Delta  \bar{P}_{ee} \equiv \bar{P}^{^7{\rm Be}}_{ee}(E_\nu,L_{max})-  \bar{P}^{^7{\rm Be}}_{ee}(E_\nu,L_{min})$$ versus $\Delta m_1^2$ and $\Delta m_2^2$ at $E_\nu=0.862$ MeV which is the energy of the $^7$Be line. As seen in these figures, for $\Delta m^2>10^{-11}$ eV$^2$, the seasonal variation can be  sizable. BOREXINO has recently published seasonal variation of the $^7$Be flux reaching the Earth. We shall examine whether this piece of information can help to constrain the parameter space of the pseudo-Dirac scheme.
\section{Analysis of the solar neutrino data \label{analysis}}  
As seen in Eqs. (\ref{PeeBULK},\ref{PesBULK}) and Eq.  (\ref{thetM}), the solar neutrino flux on the Earth depends on $\theta_{12}, \ \Delta m_{21}^2$ and $\theta_{13}$ as well as on $\Delta m_i^2$. Because of the smallness of $\sin^2\theta_{13}$,  the sensitivity of the solar data to $\theta_{13}$ is negligible so we fix  $\theta_{31} = 8.57^\circ $ throughout our analysis \cite{Esteban:2020cvm}. Historically, the solar neutrino data have provided the first measurement of $\theta_{12}$ and $\Delta m_{21}^2$ within the standard neutrino mass and mixing paradigm ({\it i.e.,} setting $\Delta m_i^2=0$). These measurements were confirmed by the KamLAND reactor experiment which measured the $\bar{\nu}_e$ flux from reactors active throughout Japan. The baseline of KamLAND, $L_{Kam}$ was less than 200~km so for the values of $\Delta m_i^2$ of our interest ($10^{-12}$ eV$^2\leq \Delta m_i^2\leq 10^{-9}$ eV$^2$), we can write $\Delta m_i^2 L_{Kam}/(2 E_\nu)\ll 1$.  Thus, the neutrino oscillation in the KamLAND experiment was sensitive only to $\theta_{12}$ and $\Delta m_{21}^2$. As a result, the determination of these parameters by KamLAND is also valid for our pseudo-Dirac scenario with $\Delta m_i^2 \ne 0 \ll 10^{-9}$ eV$^2$.
Indeed, the determination of $\Delta m_{21}^2$ by KamLAND  suffers from much smaller uncertainty than that by the solar neutrino data. We therefore treat  $\Delta m_{21}^2$  by a nuisance parameter with a mean value of $\Delta\bar{m}_{21}^2 = 7.54\times 10^{-5}$~eV$^2$ and  an error $\sigma_{\Delta m^2_{21}} = 0.5\times 10^{-5}$~eV$^2$ as measured by KamLAND  \cite{KamLAND:2010fvi}. 
The effects of $\Delta m_3^2$ on $P_{ee}$ and on $P_{es}$ are respectively suppressed by $\sin^4\theta_{13}$ and  $\sin^2\theta_{13}$ so the sensitivity to $\Delta m_3^2$ is low. We therefore {set $\Delta m_3^2 = 0$} and focus on the effects of nonzero $\Delta m_1^2$ and  $\Delta m_2^2$ on the solar neutrinos. We employ the latest solar data both from Super-Kamiokande and from BOREXINO to extract information on $\Delta m_1^2$ and $\Delta m_2^2$. We also set $\theta_{12}$ as a free parameter to ``re"-measured from the solar neutrino data in the presence of nonzero $\Delta m_1^2$ or $\Delta m_2^2$.

We use the $\chi^2$ analysis in order to constrain the allowed regions for the free parameters of the theory, separately defining  $\chi^2$ for each $\Delta m_i^2$  setting the rest equal to zero: 
\begin{align}\label{DeltaCHI}
\chi^2_{\rm min}(\Delta m_i^2,\theta_{21}) = \\ \nonumber
{\rm \min\limits_{nusiance}} & \left\lbrace \chi^2_{Su} + \chi^2_{Bo} + \left( \frac{\Delta m_{21}^2 -{ \Delta  \bar{m}_{21}^2} }{\sigma_{\Delta m^2_{21}}} \right)^2 \right\rbrace
\end{align}
where  $Su$ ($Bo$) indicates the Super-Kamiokande (BOREXINO) experiment. $\chi^2_{Su}$ is defined as  follows \cite{Super-Kamiokande:2016yck}: 
\begin{align}\label{CHi2_Su}
\chi^2_{Su}  =  \sum_{k}  \left[ \left( \frac{ D_k - f_k(\delta_B,\delta_S,\delta_R)\alpha^{B8}\mathcal{T}^{B8}_k }{\sigma_k}\right)^2 \right] \\ \nonumber
+ \left(\frac{\alpha^{B8} - 1}{\sigma_{\alpha^{B8}}}\right)^2 + \delta_B^2 + & \delta_S^2 + \delta_R^2
\end{align} 
Subscript $k$ runs over the 13 bins of $^8$B solar neutrino energy spectrum starting from $3.5 \ \rm MeV$ to $10 \ \rm MeV$. Super-Kamiokande covers the energy range $3.5 \ \rm MeV$ to $19.5 \ \rm MeV$. However, the effect of $\Delta m_i^2 $ in the range under study in this paper will be significant  only at energies lower than $10 \ \rm MeV$. As a result, we do not consider the higher energy bins and   we do not therefore  need to worry about the $hep$ data events. \footnote{Regarding to day/night effect see the Appendix.} Similarly for the BOREXINO experiment we have
\begin{equation}\label{CHi2_Bo}
\chi^2_{Bo}  =  \sum_{j}  \left[ \left( \frac{ D^j - \alpha^j \mathcal{T}^j }{\sigma^j}\right)^2  + \left(\frac{\alpha^j -1}{\sigma_{\alpha^j}}\right)^2 \right]
\end{equation} 
Superscript $j$ runs over $pp$, $^7$Be and $pep$ solar neutrino event rate (counts per day per 100 t). 

  \begin{figure}
  \centering
  	\hspace{0cm}
  	\includegraphics[width=0.5\textwidth]{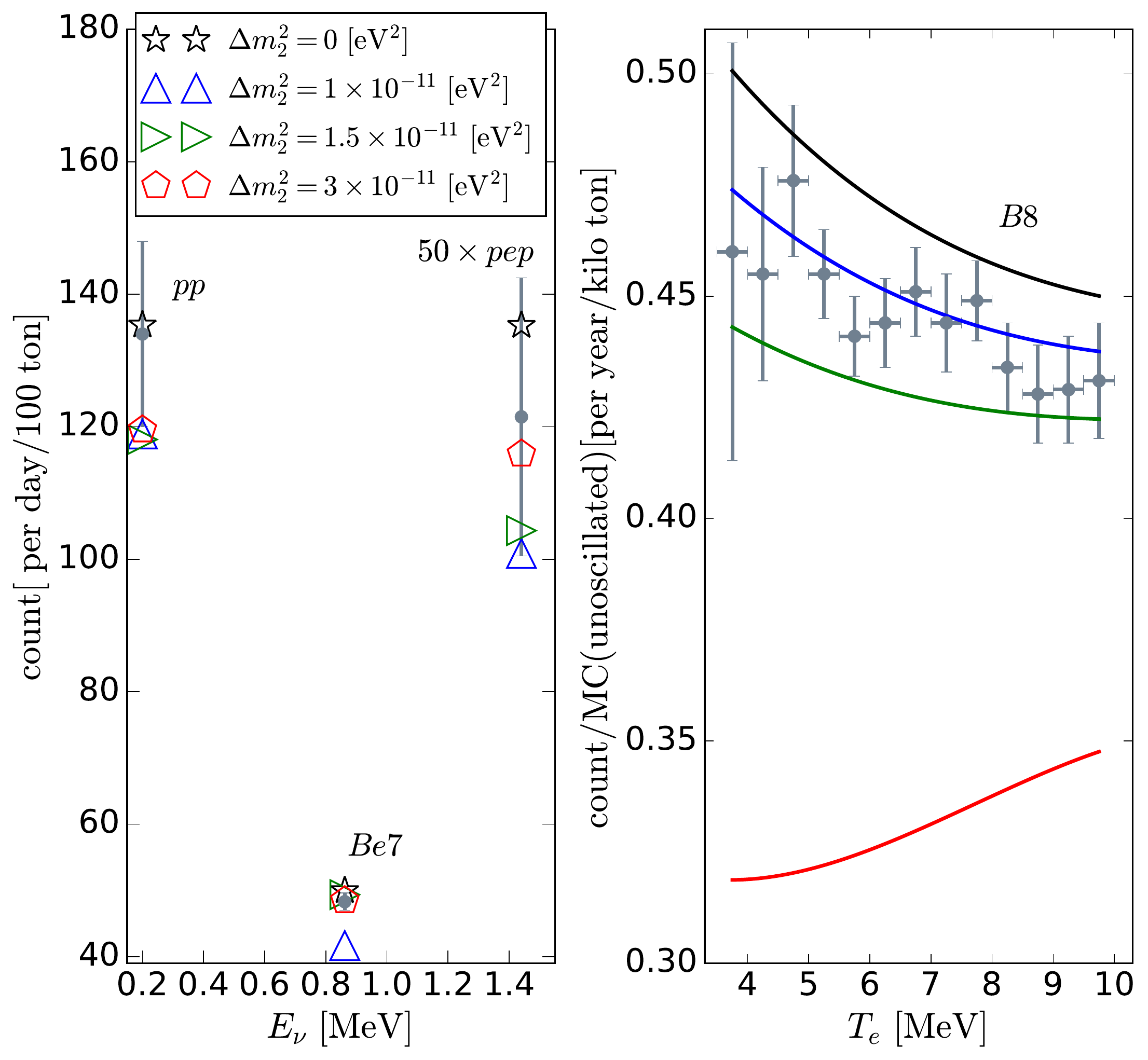}
  	\caption[...]{The annually averaged BOREXINO (left) and Super-K (right) data points. Predictions  for the splitting  values   of $\Delta m_2^2 = 1\times 10^{-11}$~eV$^2$ (blue), $\Delta m_2^2 = 1.5\times 10^{-11}$~eV$^2$ (green) and $\Delta m_2^2 = 3\times 10^{-11}$  eV$^2$ (red) are illustrated. The  standard MSW prediction (black) is added for comparison. We have taken $\Delta m_{21}^2=7.5\times 10^{-5}$~eV$^2$ and $\theta_{12}=33.4^\circ$ \cite{Esteban:2020cvm}.}
  	\label{fig:R1}
  \end{figure} 

$D$ represent the background-subtracted measured data. $\sigma $ include both statistical and systematic errors. Their values are taken from ref \cite{Super-Kamiokande:2020} and table 1 of ref \cite{BOREXINO:2020hox} for Super-kamiokande and BOREXINO respectively. While  the Super-Kamiokande data covers the solar neutrino spectrum with energies above 3.5 MeV, the BOREXINO data provides precision measurement of the low energy part of the spectrum.  The data used in Eqs. (\ref{CHi2_Su},\ref{CHi2_Bo}) is  averaged over  a year.

$\mathcal{T}$ is the prediction which will be discussed below. $\alpha^j$ are added as nuisance parameters to account for the flux normalization uncertainty in the predictions of various solar neutrino components. The uncertainty values for $pp$ and $pep$ are taken equal to $1\%$  and those for  $^7$Be and $^8$B are taken equal to  $6\%$ and $12\%$, respectively \cite{Vinyoles:2016djt}. 
We also consider the energy correlated systematic uncertainties in spectral shape, energy scale and energy resolution of $^8$B by adding nuisance parameters $\delta_B$,$\delta_S$ and $\delta_R$, respectively. The dependence of $f_k(\delta_B,\delta_S,\delta_R)$ on the nuisance parameters and energy bins are explained in ref \cite{Super-PhD}. The prediction of the theory is derived by
\begin{align}\label{eq:T}
\mathcal{T}^j_k(t_0, t_1, \Delta t) & =  \frac{t_0}{\Delta t} \mathcal{N}_{\rm det} \int_{t_1}^{t_1+\Delta t}  dt  \int dE_{\nu} \int_0^{T_e^{max}} dT_e \\ \nonumber 
R^j_k(T_e) \phi^j(L) & \frac{d\lambda^{j}}{dE_{\nu}}(E_\nu) \Biggl[ \frac{d\sigma_e}{dT_e}(E_\nu,T_e) \bar{P}^j_{ee}(E_\nu,L)  \\ \nonumber
 + \frac{ d\sigma_{\mu,\tau}}{dT_e}& (E_\nu,T_e) \biggl( 1-\bar{P}^j_{ee}(E_\nu,L)- \bar{P}^j_{es}(E_\nu,L) \biggl) \Biggl]  .
\end{align}
where $\Delta t $ is the time period over which the temporal average is taken. For annually averaged data, $\Delta t$  should be of course taken to be a year; then,  $\mathcal{T}^j_k$ will be independent of $t_1$, i.e., independent of the start of the data taking period. $t_0$ determines the temporal unit of data which for  the BOREXINO and Super-Kamiokande experiments are respectively taken to be a day and a year.  For the BOREXINO experiment,  $\mathcal{N}_{\rm det} = 3.307 \times 31 \ \rm per \ 100 \ ton$.  For the Super-Kamiokande experiment, $\mathcal{N}_{\rm det} = ({10}/{18})({1}/{m_p})$ is the number of electron at each kilo ton of the Super-Kamiokande detector and $m_p$ is  the mass of the proton in kilotons.
 $R^j(T_e)$ is the detector performance function to measure the $j$th component. $T_e$ is the recoil energy of the scattered electron. The $R^j(T_e) \sim 1$  for  all three components $(pp,pep,^7$Be) because we have used the total event rate for the case of BOREXINO. $R_{k}^{B8}(T_e)$ for $k$th energy bin of $^8B$  follows  a Gaussian function computed in Ref. \cite{Super-Kamiokande:2016yck}.

$\phi^j(L)$ is the solar neutrino flux normalization:
\begin{equation}
\phi^j(L) \equiv \bar{\phi}^j \frac{\bar{L}^2}{L^2}
\end{equation}
$\bar{\phi}^j$ is solar standard model prediction \cite{Vinyoles:2016djt} calculated at $\bar{L}$, temporal average of sun to earth distance. $L$ is the  Sun to Earth distance  which varies during a  year due to the Earth orbit eccentricity. $\frac{d\lambda^j}{dE_{\nu}}(E_\nu)$ is the solar neutrino spectrum.  For the $pp$ and $^8B$ components which have continuous  spectrum, $\phi^{pp}(L) ({d\lambda^{pp}}/{dE_{\nu}})$ and $\phi^{B8}(L) ({d\lambda^{B8}}/{dE_{\nu}})$ are in unit of $[\rm cm^{-2} s^{-1} MeV^{-1}]$.  The normalization of the monoenergetic  $^7$Be and $pep$ fluxes   are in  the unit of $[\rm cm^{-2} s^{-1}] $.

The differential cross sections of the electron scattered by neutrinos of different flavors ($e,\mu,\tau$) are \cite{Chen:2021uuw}:
\begin{align}
\frac{d \sigma_{e (\mu ,\tau)}}{d T_e} =  \\ \nonumber 
\frac{G_F^2 m_e}{2\pi}& \Biggl[  \biggl( 2\sin^2(\theta_W)  \pm 1 \biggl)^2 + \biggl( 2\sin^2(\theta_W)(1-\frac{T_e}{E_\nu}) \biggl)^2 \\ \nonumber
& -2\sin^2(\theta_W)  \biggl( 2\sin^2(\theta_W) \pm 1 \biggl) \frac{m_e T_e}{E_\nu^2} \Biggl]
\end{align}
where for $\nu_e$  (for $\nu_\mu,\nu_\tau$) we should take the plus (minus) sign.  We take the Weinberg angle as $\sin^2(\theta_W) = 0.22342$\footnote{\url{https://pdg.lbl.gov/}}. The maximum recoil energy of the electron is given by $$T_e^{max} = \frac{E^2_\nu}{E_\nu + m_e/2}.$$
\section{Results\label{results}}
In sect. \ref{integrated},  we first study how by combining the BOREXINO data on the $^7$Be, $pep$ and $pp$  event rate with the Super-Kamiokande solar neutrino spectrum data, we can constrain $\Delta m_1^2$ and  $\Delta m_2^2$. Surprisingly, we find that there is a new solution in the range of $\Delta m_2^2 = (1-2)\times 10^{-11}$~eV$^2$. We discuss whether the measurement of the total active neutrino fluxes by SNO or by current and future direct dark matter search experiments can test this new solution with nonzero $\Delta m_2^2$. In sect. \ref{seasonal}, we  study the effects of  $\Delta m_1^2$ and $\Delta m_2^2$  on the seasonal variation of the $^7$Be flux and contrast it with the recent BOREXINO data release on the seasonal variation of  the $^7$Be  flux on the Earth.  Surprisingly, this data independently points towards the same  solution. We show that our new solution  favors the GNO radio chemical experiment  and is compatible with combined experiments Gallex and GNO \cite{GNO:2005bds}. We then discuss the prospect of testing this solution by a more precise measurement of  the seasonal variation of the $^7$Be flux.
\subsection{Total solar flux integrated over year(s) \label{integrated}}
In this sub-section, we  analyze the time-integrated BOREXINO and Super-Kamiokande solar neutrino data.
The data points are shown in Fig.~\ref{fig:R1}. The vertical axis in the left (right) panel is the number of counts per day per 100 tons (the number of counts over MC (unoscillated) per year per kilo ton). The prediction of the pseudo-Dirac scenario with $\Delta m_2^2 = (1,1.5,3)\times 10^{-11}$ eV$^2$ are also shown. To obtain these predictions, we have set $\Delta m_1^2 =0$ and have used  Eq.~(\ref{eq:T}). The standard MSW scheme ({\it i.e,}  $\Delta m_1^2=\Delta m_2^2 =0$) is added for  comparison.  As seen from the figure, large values of the splitting such as $\Delta m_2^2 \sim 3\times 10^{-11}$  eV$^2$  can be  ruled out by the $^8B$ data points.  
Although the range few$\times 10^{-12}$~eV$^2 <\Delta m_2^2<10^{-11}$ eV$^2$ is consistent with the Super-Kamiokande data, it is located out of one sigma error of the precise $^7$Be line measurement by BOREXINO. As demonstrated by the green curve and triangle, The $\Delta m_2^2 = 1.5\times 10^{-11}$ eV$^2$ also gives a good fit to $^8$B data points as well as to the BOREXINO data points. 
Ref. \cite{Anamiati:2017rxw} had also found this solution with $\Delta m_2^2 \sim 10^{-11}$ eV$^2$. Our results with updated solar data \cite{BOREXINO:2020hox} which includes the relatively precise pep line measurement confirms their finding.
Notice that the prediction with $\Delta m_2^2=1.5 \times 10^{-11}$ eV$^2$ for the $pep$ line is smaller than that with $\Delta m_2^2=0$. Improving the precision of the $pep$ line can therefore test this non-trivial solution.
  \begin{figure}
  \centering
  	\hspace{0cm}
  	\includegraphics[width=0.45\textwidth]{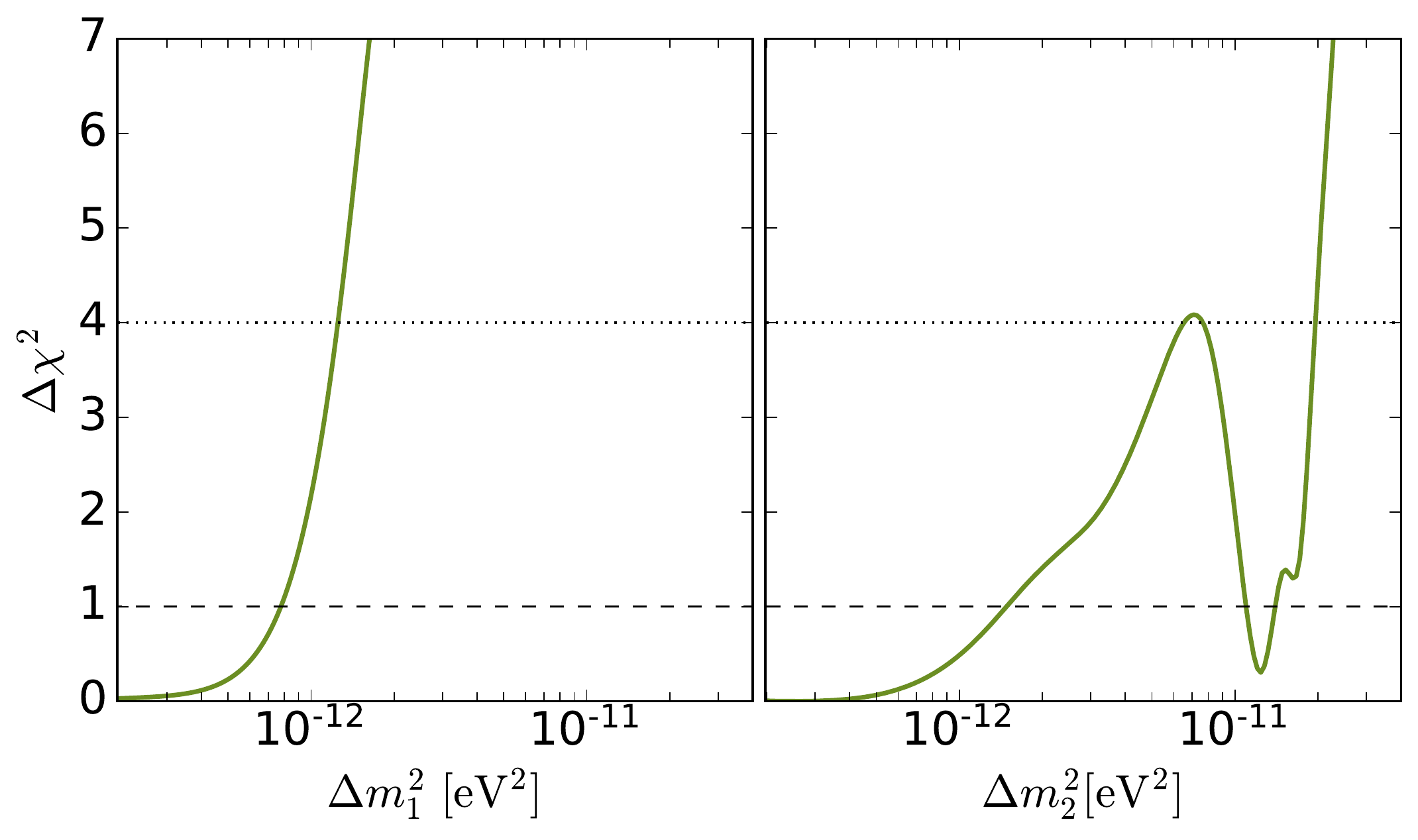}
  	\caption[...]{$\Delta \chi^2$ as a function of $\Delta m_i^2$ minimized over $\theta_{12}$. The dashed and dotted horizontal lines respectively correspond to  the $68\%$ and $95.45\%$ confidence levels.}
  	\label{fig:R2}
  \end{figure}  
  
\begin{figure}
    \centering
  	\hspace{0cm}
  	\includegraphics[width=0.45\textwidth]{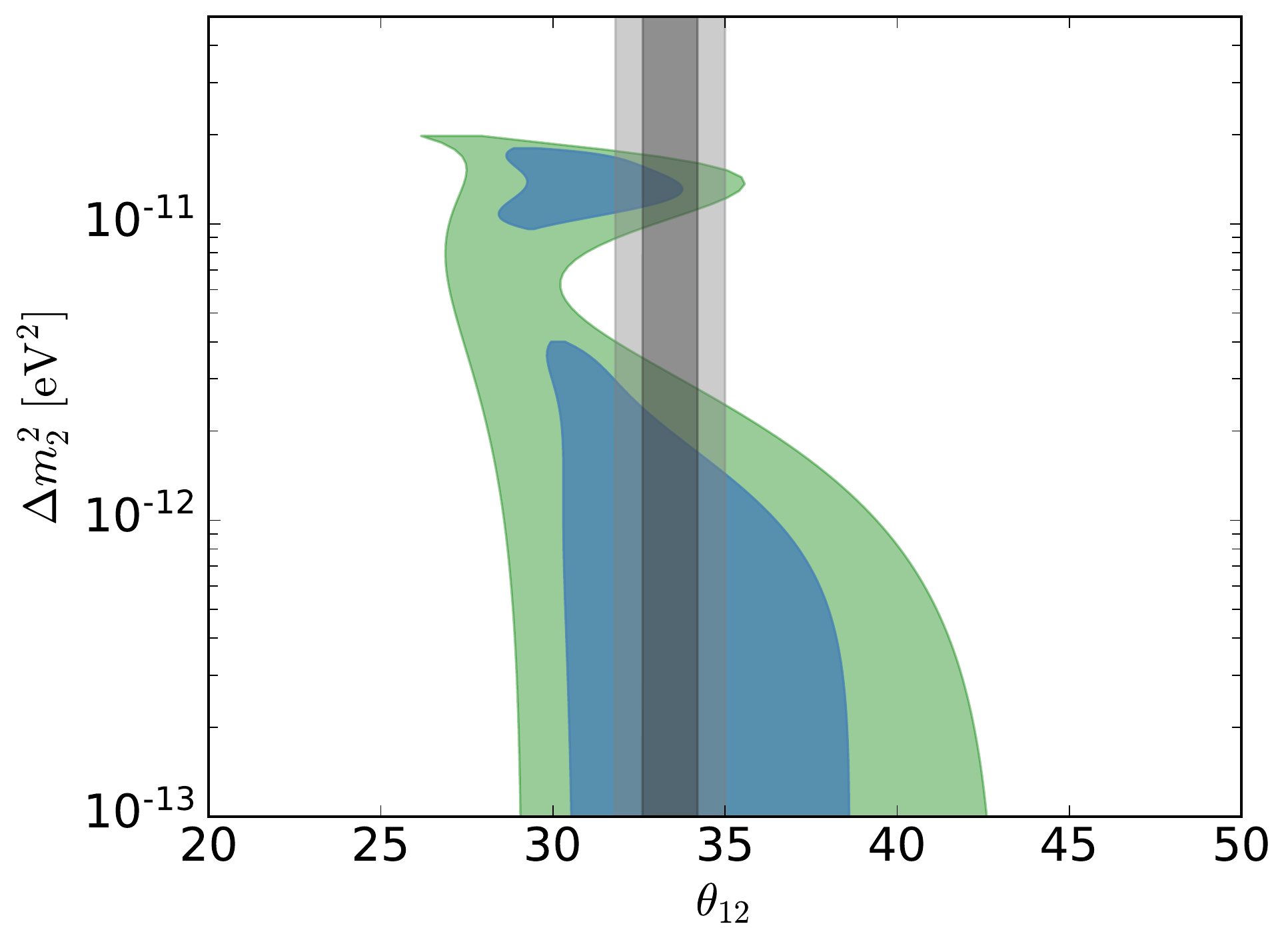}
  	\caption[...]{Allowed $1\sigma$ and $2\sigma$ regions of joint $\chi^2$ analysis for the $(\Delta m^2_2,\theta_{21})$ space, assuming  $\Delta m^2_1$  is zero. The black and grey band show the $1\sigma$ and $2\sigma$ allowed range of $\theta_{12}$ by global neutrino data analysis \cite{Esteban:2020cvm}.}
  	\label{fig:R3}
  \end{figure}

Fig. \ref{fig:R2} shows $\Delta \chi^2$ versus $\Delta m_1^2$ and $\Delta m_2^2$. $\chi^2$ is defined by  Eq.~(\ref{DeltaCHI}). To compute  $\Delta \chi^2$, we have minimized over $\theta_{12}$ and have subtracted the minimum $\chi^2$ with respect to $\Delta m_i^2$ . As seen from the figure, the values of $\Delta m_2^2$ ($\Delta m_1^2$) larger than $2\times 10^{-11}$ eV$^2$ ($1.5\times 10^{-12}$ eV$^2$) is ruled out at 2$\sigma$. This figure also demonstrates that $1\times 10^{-11}~eV^2<\Delta m_2^2<2\times 10^{-11}~eV^2$ provides a fit comparable with SM when $\Delta m_2^2\to 0$.{\footnote{We have also performed a similar analysis with the official  Super-Kamiokande data release in 2016 \cite{Super-Kamiokande:2016yck} which confirmed the current results, except that those data tended to have lower values and thus the region found in the interval of $\Delta m_{2}^2 = (1,2)\times 10^{-11}$ compared to the $\Delta m_{2}^2 \to 0$ have lower $\chi^2$ value; {\it i.e.,} providing a slightly better fit than the standard MSW with  $\Delta m_{2}^2 =0$.}}  $\Delta m_2^2<1.5\times 10^{-12}$ eV$^2$ is allowed within 1$\sigma$ C.L.  The 1$\sigma$ and  2$\sigma$ 
contours  of $\Delta m_2^2$ versus $\theta_{12}$ are shown in Fig. \ref{fig:R3} . As seen from the figure, the values of $\theta_{12}$ at the solutions that we have found are consistent with the $\theta_{12}$ measurement by the global neutrino data analysis.

 Let us now discuss the implication of the SNO  measurement of the total active solar neutrino flux. The SNO experiment has extracted the total flux by measuring the Deuteron dissociation rate, $\nu+D \to \nu+n+p$ with a precision of $8\%$  \cite{SNO:2011hxd,SNO:2011ajh}.
 This measurement is well-consistent with the standard solar model prediction for the  total neutrino flux within the uncertainties. In our model, the measured total active flux is suppressed by $(1-P_{es}(E_\nu))$. The SNO detection threshold is practically above 5~MeV.\footnote{ The natural energy threshold, which is set by the binding energy of the  Deuteron nucleus, is  2.2 MeV.}  For $E_\nu>5$~MeV and $\Delta m_2^2<2\times 10^{-11}$ eV$^2$, $P_{es}$ is below 10\% and as a result, the suppression of the total active flux measurement relative to the SM prediction will be within the flux prediction uncertainty of 12 \% \cite{Vinyoles:2016djt} and cannot therefore be resolved. For lower energies (below the SNO threshold), the deviation should be more significant. The total flux with lower energy threshold can also be measured by the coherent elastic neutrino nucleus scattering at large scale direct dark matter experiments such as the ongoing XENONnT and LZ experiments and future DARWIN experiment, promising to test this model.

\subsection{Seasonal variation\label{seasonal}} 
    \begin{figure*}
    \centering
	\hspace{0cm}
	\includegraphics[width=1.\textwidth]{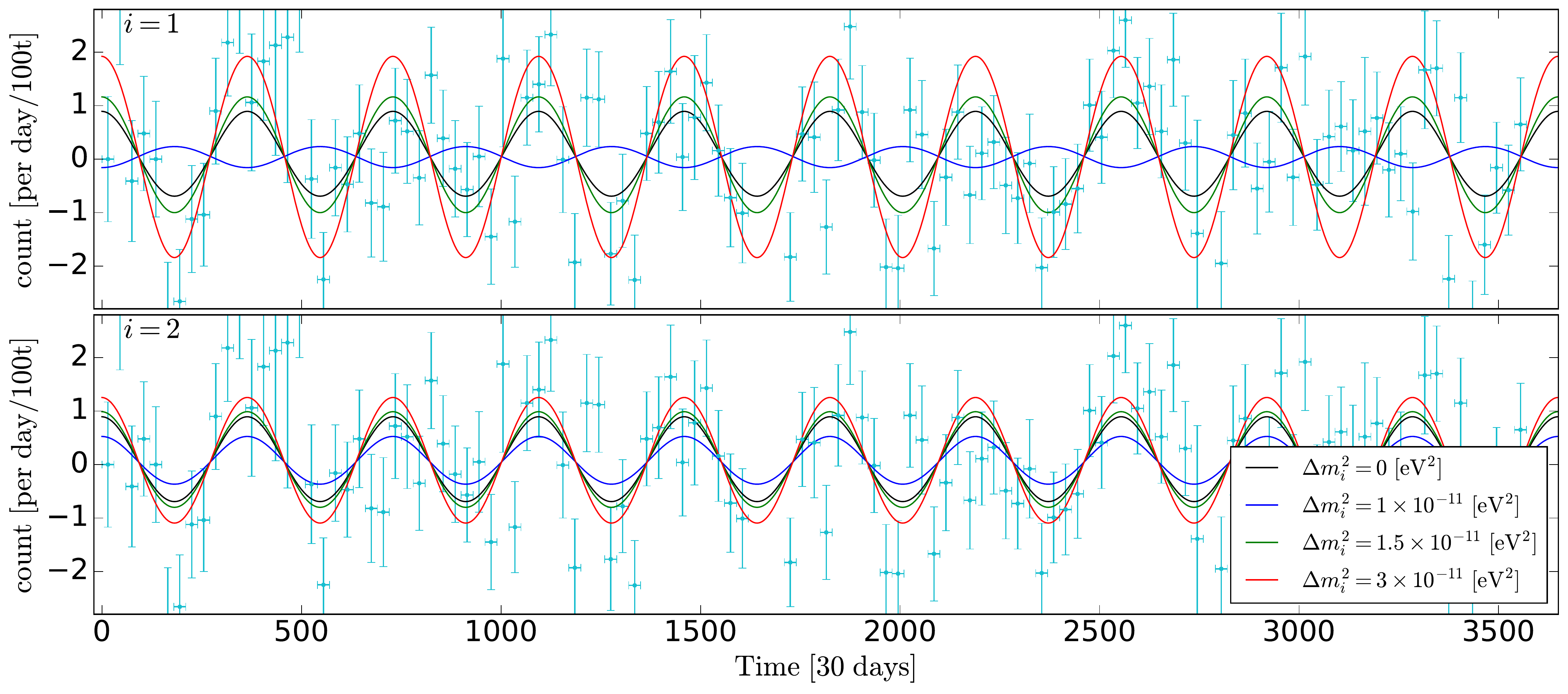}
	\caption[...]{Seasonal variation of event rate (per day per 100t) for $^7$Be.  The data points are taken from \cite{BOREXINO:2022wuy}.   The predictions for various values of $\Delta m_i^2$ are shown by curves. The  upper (lower) panel corresponds to nonzero $\Delta m_1^2$ (nonzero $\Delta m_2^2$).  The values of the standard mixing parameters are fixed to $\theta_{12}=33.4^\circ$ and $\Delta m_{21}^2=7.5\times 10^{-5}$ eV$^2$.
		\label{fig:R4} }
\end{figure*}
Recently, the BOREXINO experiment has released data on the residual of the $^7$Be neutrino event rate, showing modulation due to the seasonal variation \cite{BOREXINO:2022wuy}.  The selected events include an electron with a recoil energy larger than $0.3$ MeV. The rate is given in terms of  per day per 100t. The data points, covering a period of almost 10 years, are a time series binned in time intervals of 30 days. The annual trend of the data is subtracted.\footnote{For the exact definition of tend and residue, the readers may consult \cite{BOREXINO:2022wuy}. } 
We use this new data to independently  examine the validity of   the new solution ($\Delta m_2^2\simeq (1-2)\times 10^{-11}$~eV$^2$) found in sect. \ref{integrated}. Furthermore, studying the seasonal variation  
is an alternative approach to probe the pseudo-Dirac mass splitting. The data points along with the predictions with various values of $\Delta m_1^2$ and $\Delta m_2^2$ are shown in  Fig.~\ref{fig:R4}.

\begin{figure}
     \centering
  	\hspace{0cm}
  	\includegraphics[width=0.45\textwidth]{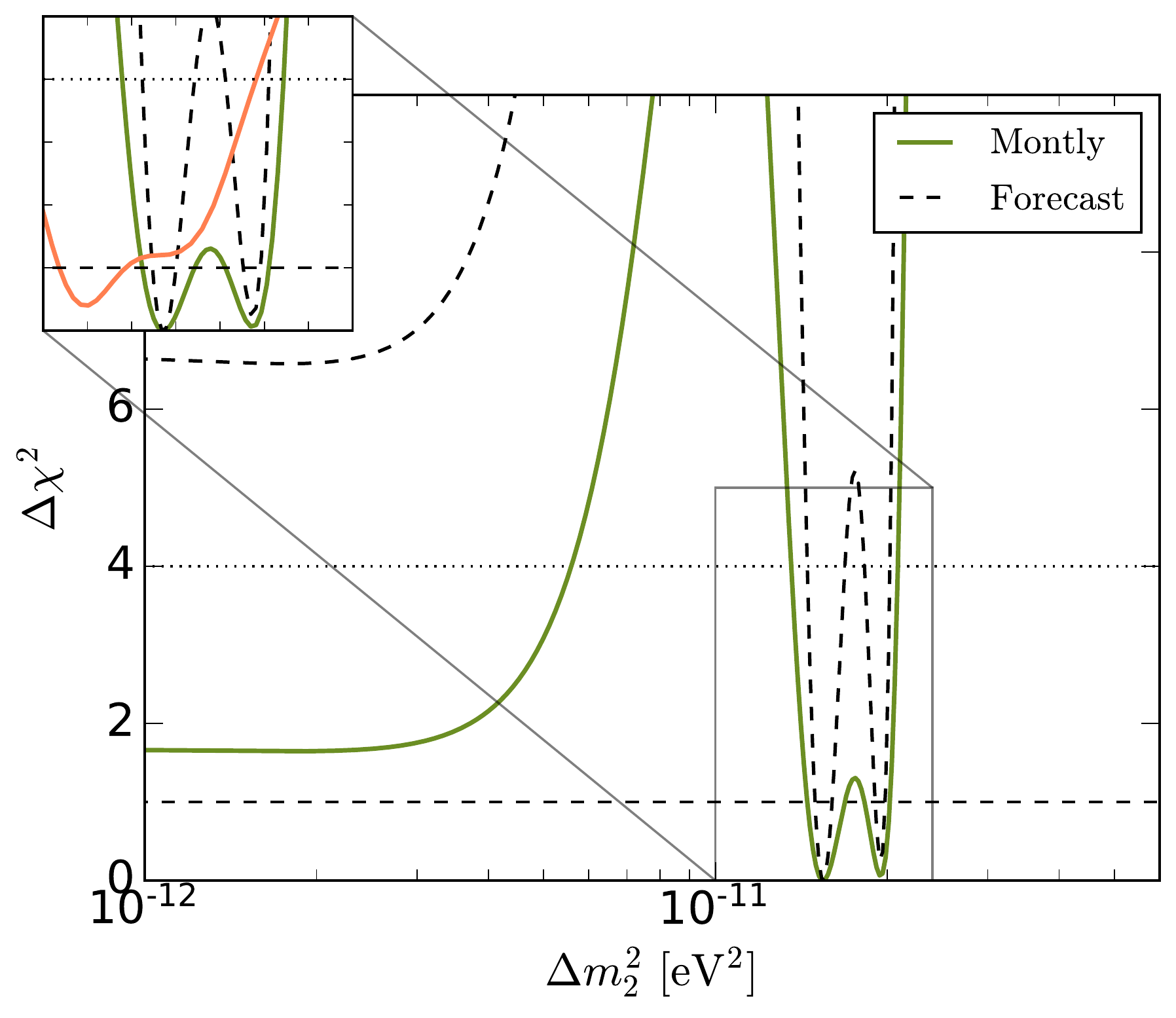}
  	\caption[...]{$\Delta \chi^2$ as a function of $\Delta m_2^2$ using just the seasonal variation data set. The values of the standard mixing parameters are fixed to $\theta_{12}=33.4^\circ$ and $\Delta m_{21}^2=7.54\times 10^{-5}$ eV$^2$.  The  dashed curve illustrates the forecast for similar experiment with  error $\sigma_t\sim 0.5$ per day per 100t; {\it i.e.,} with errors reduced  to  almost half of the current values.  The right panel of Fig.~\ref{fig:R2} is added as orange curve for comparison in the  zoomed box.}
  	\label{fig:R5}
  \end{figure}

As seen in Fig. \ref{shaky}, depending on the exact value of   $\Delta m_1^2$ and/or $\Delta m_2^2$  in the range $\sim (1-2) \times 10^{-11}$~eV$^2$, the pseudo-Dirac scheme can lead to enhancement or suppression of the seasonal   modulation. This behavior is also confirmed in Fig.~\ref{fig:R4}. In the following, we focus on $\Delta m^2_2$  and set $\Delta m^2_1 = 0 $. To constrain  $\Delta m^2_2$, we define $\chi^2$ as
\begin{equation}
\chi^2 = \sum_t \frac{\left[ D^{Be}_t - \biggl( \mathcal{T}^{Be}({\rm day}, t,{\rm m}) - \mathcal{T}^{Be} ({\rm day, t, y}) \biggl) \right]^2}{\sigma^2_{t}}
\end{equation} 
where $t$ runs over the 120 bins, each bin corresponding to a one month data taking periods. $\rm m$ and $\rm y$ stand for month and year respectively. $D^{Be}_t $ is residual of the events per day per 100 ton which are modeled as a time series trend \cite{BOREXINO:2022wuy}. $\sigma^2_{t}$ is the corresponding error at each bin $t$, as shown in Fig. \ref{fig:R4}.  $ \mathcal{T}^{Be}({\rm day}, t, {\rm month})$ is computed using Eq.~(\ref{eq:T}) by replacing the lower limit of the electron recoil energy with $T_e = 0.3 $ MeV and integrating over monthly periods.  $ \mathcal{T}^{Be}({\rm day}, t, {\rm year})$ is computed using the  same formula with an averaging priod of a year. As discussed before, $ \mathcal{T}^{Be}({\rm day}, t, {\rm year})$ should be independent of $t$.  We fix $\theta_{12}=33.4^\circ$, $\Delta m_{21}^2=7.54\times 10^{-5}$ eV$^2$  and  $\Delta m_{1}^2=0$ but vary  $\Delta m_2^2$.  Similarly to the previous section, we invoke the  standard solar model for the flux normalization, with negligible uncertainty.
As seen from Fig.~\ref{fig:R5}, the $\chi^2$ analysis using this new data set independently supports the enhancement of the modulation which occur in the  $1.4\times 10^{-11}~{\rm eV}^2 < \Delta m_2^2 < 2\times 10^{-11}~{\rm eV}^2$. This non-trivial solution falls  in the 2$\sigma$ region of the annually averaged data that we  have found in sect. \ref{integrated}.  The non-trivial solution that we have found provides a better fit to  the seasonal variation (see Fig.~\ref{fig:R5}) than the standard model with $\Delta m_2^2 \to 0$.  This is because the data shows  about $ 10 \%$ more enhanced  modulation than the $1/L^2$ modulation expected in the standard MSW scenario. On the other hand, a cancellation  on the modulation  takes place in the range $0.6\times 10^{-11}~{\rm eV}^2 < \Delta m_2^2 < 1.3\times 10^{-11}~{\rm eV}^2$ which is clearly ruled out with current data.
The standard solution  ($\Delta m_2^2 \to 0$) is allowed at just 80 \% confidence level. However, the fact that the two independent measurements, namely the Super-Kamiokande time integrated solar data with $E_\nu> 3$ MeV and the seasonal variation of the $^7$Be flux measured by BOREXINO, as well as the time integrated BOREXINO solar neutrino data simultaneously  point towards the same nonzero value of $\Delta m_2^2$ makes it imperative to look for ways to test this new solution. 

\begin{figure}
     \centering
  	\hspace{0cm}
  	\includegraphics[width=0.45\textwidth]{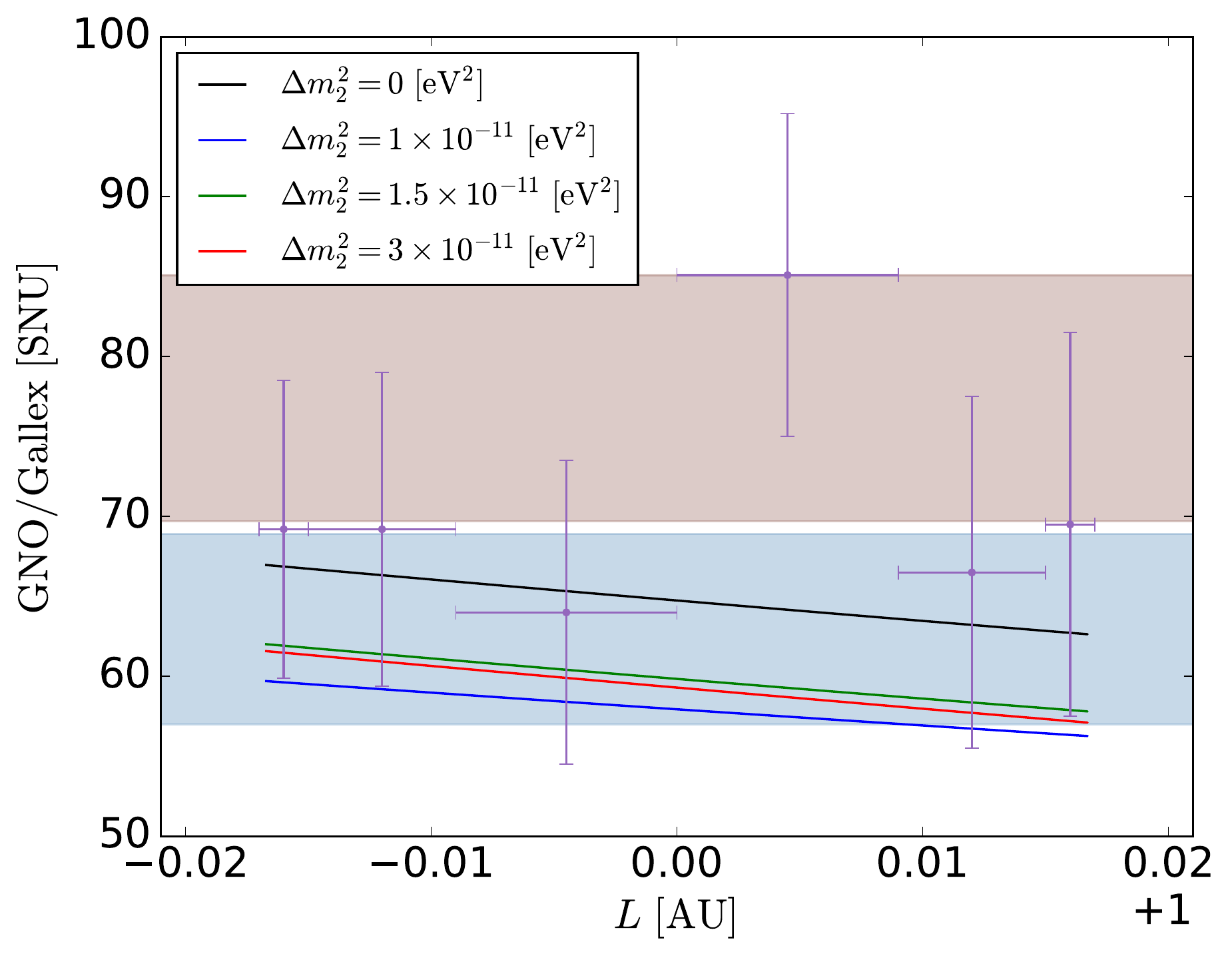}
  	\caption[...]{Combined Gallex and GNO in solar neutrino unit (SNU) versus $L$. The color code for different $\Delta m_2^2$ is same as previous plots. The brown (blue) band is the  1$\sigma$ allowed region  derived by Gallex (GNO),  averaging over time.}
  	\label{fig:gno}
  \end{figure}

We also examine this non-trivial solution with the data previously released by the radio chemical gallium experiments Gallex and GNO. We use the combined result for the variation of the total $\nu_e$ during a year. The data is taken from fig 7 of \cite{GNO:2005bds}. The results are illustrated in Fig.~\ref{fig:gno}. These two experiments had capability to record the low energy  $\nu_e$  above a threshold of $0.24 \ \rm MeV$. They might  therefore be used to examine the low energy effect of pseudo-Dirac scheme for both annually averaged along with seasonal variation data. Although, the region of our interest, $\Delta m_2^2 \sim 1.5 \times 10^{-11} \rm eV^2$, is compatible with  the annually averaged data, due to large uncertainty the effect of seasonal variation is not apparent. The main  source of  uncertainties comes from the deviation of Gallex results from  the GNO data.  We have highlighted the 1$\sigma$ averaged result of the Gallex  in the Fig.~\ref{fig:gno}  with light brown. It is clear that the combined result is affected by the tendency of Gallex data to higher values which is even at odds  to the standard MSW scheme. We have also highlighted the 1$\sigma$ averaged result of GNO with light blue. There is no preferences between the non-trivial solution and standard MSW when we just consider the GNO result at annually averaged level.

In the previous sections, we proposed four alternative  methods to test this solution. In the following, we investigate how by improving the seasonal variation measurements, this new solution can be tested.

  Let us suppose the true value of $\Delta m_2^2$ is close to the best fit that we have found and then study what level of precision is required to rule out the standard solution with  $\Delta m_2^2=0$.
  The dashed line in Fig.~\ref{fig:R5} shows the value of $\Delta \chi^2$ versus $\Delta m_2^2$ setting $\sigma_{t} \sim 0.5 \ (\rm count / day \ per \ 100t)$. Notice that such $\sigma_t$ requires a factor of 2 reduction in the current uncertainty. As seen from the figure, with such an improvement, the standard solution can be ruled out at better than 2 $\sigma$ C.L.
  
\begin{figure}
      \centering
  	\hspace{0cm}
  	\includegraphics[width=0.45\textwidth]{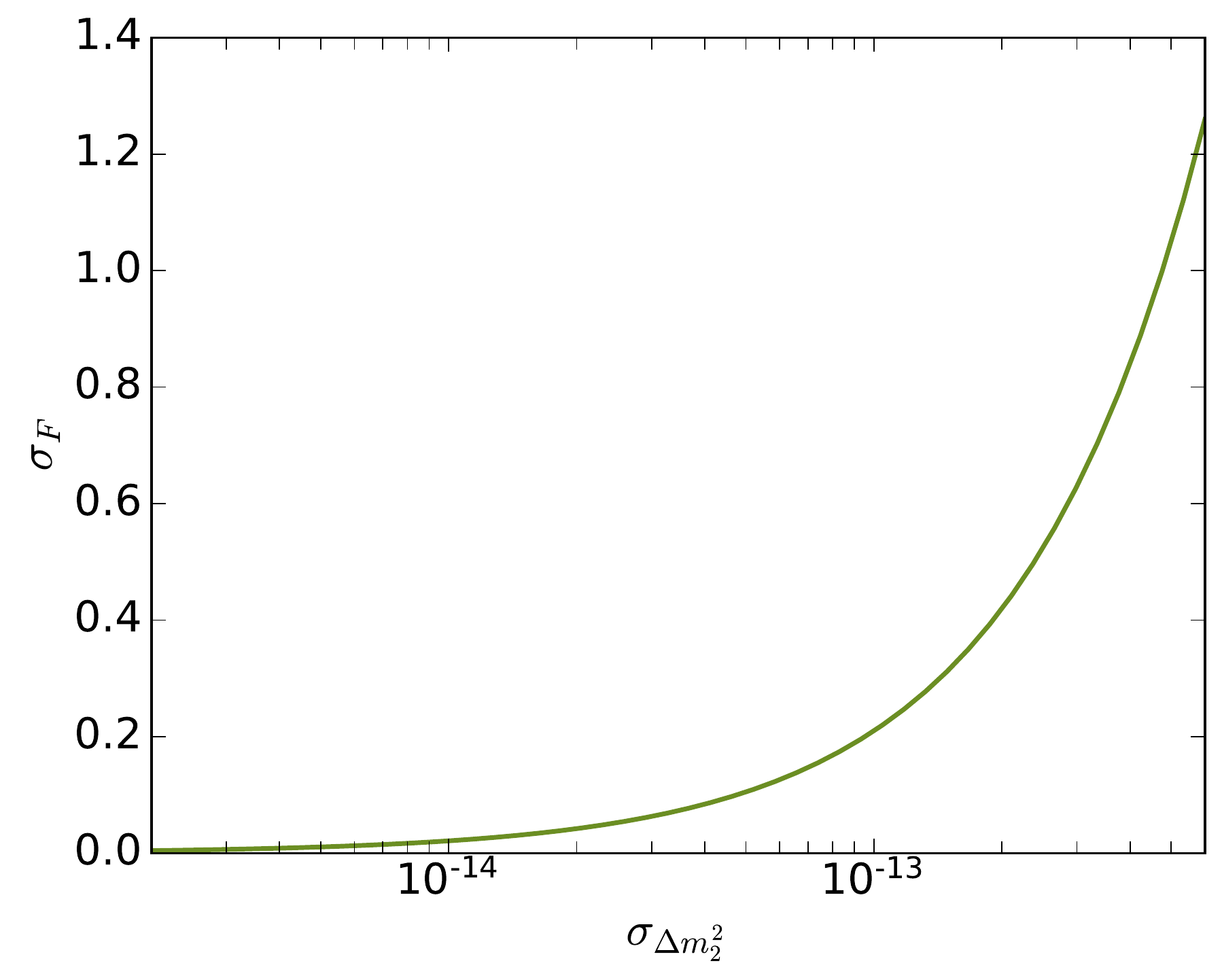}
  	\caption[...]{The precision required in the time variation measurement to  determine $\Delta m_2^2$ with a  precision of $\sigma_{\Delta m_2^2}$. We  have assumed that the true value is  $\Delta m_2^2 = 1.5\times 10^{-11}$  eV$^2$.}
  	\label{fig:R6}
  \end{figure}

Let us now discuss how small  $\sigma_t$ should be in order  to obtain a desired precision on  $\Delta m_2^2$. To answer this question, we assume  that the error value $\sigma_t \sim \sigma_F$ is equal for all bins and utilize the Fisher forecast formalism \cite{Tegmark:1996bz} with 
 \begin{equation}
 	\sigma_F = \sigma_{\Delta m^2_i} \sqrt{\sum_t \left( \frac{\partial \mathcal{T}^{Be}({\rm day},t,{\rm m})}{\partial \Delta m^2_i} \right)^2}
 \end{equation}
$\sigma_{F}$ is the ideal measurement error in order to have $2\times \sigma_{\Delta m^2_i}$, the $1\sigma$ allowed region for parameter $\Delta m^2_i$.  The sum is over one year data points binned in 30 days and $\mathcal{T}^{Be}$ is the prediction for $^7$Be neutrino event rate with $T_e > 0.3 \ {\rm MeV}$, similarly to the current measurement. We assume the true value $\Delta m_2^2 = 1.5\times 10^{-11}$ eV$^2$ which is in the range of $\sim (1-2)\times 10^{-11}$  eV$^2$. The result is shown in Fig.~\ref{fig:R6}.  Error values of  order of $\sigma_{F} \sim 1 \ (\rm count / day 100t)$ lead to $1\sigma$ range $2\times \sigma_{\Delta m^2_i} \sim 10^{-12}$  eV$^2$. In particular, we obtain  $2\times \sigma_{\Delta m^2_i} \sim 0.5 \times 10^{-12}$  eV$^2$ reducing them to   $\sigma_{F} \sim 0.5 \ (\rm count / day 100t)$.

 \section{Discussion and conclusions\label{Dis}}
We have studied the oscillation of the solar neutrinos within the pseudo-Dirac  scheme. Our focus has been on the splittings of $\nu_1$ and $\nu_2$ states, $\Delta m_1^2$ and $\Delta m_2^2$ of order of $10^{-13}$ eV$^2-10^{-10}$ eV$^2$ which are relevant for solar neutrinos. Since the contribution of $\nu_3$ to the solar neutrinos ($\nu_e$ at the  production) is suppressed by $\theta_{13}$, a splitting in $\nu_3$ will not affect the solar neutrino data. To derive bounds on the splitting, we have used the latest BOREXINO and Super-Kamiokande solar neutrino data  and have employed the $\Delta m_{21}^2$ measurement by KamLAND. We have found that these data rule out $\Delta m_1^2$ and $\Delta m_2^2$ above $2\times 10^{-11}$ eV$^2$.  However, we find a new solution in the range of $\Delta m_2^2\sim (1-2) \times 10^{-11}$ eV$^2$ \  and  $\Delta m_1^2=0$ which fits the solar neutrino data (especially the $^8$B data measured by Super-Kamiokande)  in addition to  the standard three neutrino scenario with  $\Delta m_1^2=\Delta m_2^2=0$. We have discussed the possibility of ruling out this solution with the total active neutrino flux measurement by SNO. We found that the deviation due to $P(\nu_e \to \nu_s)$ at this solution for neutrinos with energy above the SNO detection threshold can hide within the flux prediction. We have examined the robustness of this new solution against the accumulation of more solar data. The data available by 2016 slightly prefers this solution to the standard MSW. Ref. \cite{Anamiati:2017rxw} also confirms this solution.

We have examined the possibility of testing this non-trivial pseudo-Dirac solution with the recent data release by BOREXINO on the seasonal variation of $^7$Be  \cite{BOREXINO:2022wuy}. 
Surprisingly, the seasonal variation also points towards a solution with $1.4 \times 10^{-11}{\rm eV}^2<\Delta m_2^2<2\times 10^{-11}{\rm eV}^2$, independently.  Indeed, this solution fits the seasonal variation better than the standard three neutrino scheme but, the $\Delta m_1^2=\Delta m_2^2=0$ solution is still allowed at 80\% C.L. We have discussed how reducing the uncertainty in the measurement of seasonal variation can help to measure $\Delta m_2^2$ with better precision or set a bound on it. 

We also examine our new solution with radio chemical gallium experiments Gallex and GNO. The solution is in agreement with the combined Gallex and GNO averaged results, favoring GNO over Gallex. However, due to large uncertainties it is not possible to test the effect of seasonal variation with this old data.

We have proposed four independent approaches to test the non-trivial solution that we have found: (1)
Measurement of the $^8$B  flux in the energy range below 4 MeV  with a moderate precision of 20 \%  (or better) can test the solution. The proposed THEIA detector \cite{Theia:2019non}, with a relatively low detection energy threshold will be able to perform such a measurement.
 (2)  By improving the precision of the measurement of the $pep$ line, the solution can be tested. (3)  Reducing the uncertainty in the seasonal variation of the $^7$Be line by half can test this solution. (4)  Finally, the measurement of the total active solar flux via coherent elastic $\nu$ nucleus scattering by direct dark matter search experiments can provide an alternative method for testing the
 solution.
 
We have found that both time integrated solar neutrino data and the $^7$Be time variation, independently from each other,  constrain $\Delta m_1^2<1.5 \times 10^{-12}$ eV$^2$ and $\Delta m_2^2<2\times 10^{-11}$ eV$^2$ at $2 \sigma$ C.L.

All material and code of this article are publicly available on \href{https://github.com/SaeedAnsarifard/SolarNeutrinos-pseudoDirac.git}{https://github.com/SaeedAnsarifard/SolarNeutrinos-pseudoDirac.git} 
\appendix*

\section{Pseudo-Dirac scheme in the presence of matter}
In this appendix, we derive dispersion relation and the energy momentum eigenstates for the pseudo-Dirac scheme in the presence of matter effects. We compute $P(\nu_e\to \nu_e)$ and $P(\nu_e\to \nu_s)$ for solar neutrinos in the pseudo-Dirac neutrino scheme. We then formulate the time dependence (seasonal variation) of the flux arriving to the Earth, considering the eccentricity of the Earth orbit around the Sun.

Let us start with one flavor state with the effective Lagrangian
\begin{widetext}
\begin{eqnarray}
\mathcal{L} &=& \bar{\Psi} i \partial \cdot \gamma \Psi -m  \bar{\Psi} \Psi -V   \bar{\Psi}  \gamma^0 P_L \Psi -\mu  \bar{\Psi}^cP_R \Psi  -\mu  \bar{\Psi}P_L\Psi^c \cr &=&
\bar{\Psi}^c i \partial \cdot \gamma \Psi^c -m  \bar{\Psi}^c \Psi^c +V   \bar{\Psi}^c  \gamma^0 P_R \Psi^c -\mu  \bar{\Psi}^c P_R \Psi  -\mu \bar{\Psi}P_L\Psi^c
\label{LAG}
\end{eqnarray}
\end{widetext}
where $\Psi$ is a general Dirac spinor and $\Psi^c =-i\gamma^2\Psi^*$.  Taking the derivative  of the Lagrangian in the first line of  Eq. (\ref{LAG}) with respect to $\bar{\Psi}$, we arrive at the  
Euler-Lagrange equation,
\be i \partial \cdot \gamma \Psi-m\Psi-V \gamma^0 P_L \Psi -\mu P_L \Psi^c=0 . \label{Euler1}\ee
Similarly  taking derivative  of the Lagrangian in the second line of  Eq. (\ref{LAG}) with respect to $\bar{\Psi}^c$, we find
\be i \partial \cdot \gamma \Psi^c-m\Psi^c+V \gamma^0 P_R \Psi^c -\mu P_R \Psi=0 . \label{Euler2}\ee
Applying $(i\gamma\cdot \partial-V\gamma^0P_L)$ and  $(i\gamma\cdot \partial+V\gamma^0P_R)$  respectly to Eqs. (\ref{Euler1},\ref{Euler2}), we obtain the following relations
\begin{align}
	-\partial^2 \Psi-2iVP_L\partial_3 \Psi & = m^2 \Psi +m\mu\Psi^c-\mu^2 P_R\Psi \\ \nonumber 
	-\partial^2 \Psi^c+2iVP_R\partial_3 \Psi^c & = m^2 \Psi^c +m\mu\Psi +\mu^2 P_L\Psi^c 
\end{align}
where we have taken the third ($z$) direction along the momentum  ($p$) of the particle.
Remembering $P_L \Psi=\nu_L$, $P_L \Psi^c=\nu_R^c$, $P_R \Psi=\nu_R$, $P_R \Psi^c=-\nu_L^c$ and using $P_LP_R=0$, we obtain
\begin{align}\label{nuLR}
 (E^2-p^2)\left( \begin{matrix} \nu_L \cr \nu_R^c \end{matrix}\right) = & \left( \begin{matrix} 2pV+m^2 & m\mu \cr m\mu & m^2+\mu^2 \end{matrix}\right) \left( \begin{matrix} \nu_L \cr \nu_R^c\end{matrix} \right) \\ \nonumber
 =&\left[\left( \begin{matrix} 2pV &0 \cr 0 & 0 \end{matrix}\right) +\left( \begin{matrix} 0 &m \cr m & \mu \end{matrix}\right)^2 \right] \left( \begin{matrix} \nu_L \cr \nu_R^c\end{matrix} \right)
\end{align}
and
\be (E^2-p^2)\left( \begin{matrix} \nu_R \cr \nu_L^c \end{matrix}\right)=\left( \begin{matrix} m^2 -\mu^2& -m\mu \cr -m\mu & m^2-2pV \end{matrix}\right) \left( \begin{matrix} \nu_R\cr \nu_L^c\end{matrix} \right)\ee
Let us focus on Eq.~(\ref{nuLR}). We should of course take the ultra relativistic limit, $p\gg V,m,\mu$ so the energy eigenvectors correspond to the eigenvectors of 
\be \left( \begin{matrix} V+\frac{m^2}{2p} & \frac{m\mu}{2p}  \cr  \frac{m\mu}{2p}  &  \frac{m^2+\mu^2}{2p} \end{matrix}\right).\ee
In the limit $2Vp\ll m\mu$, we  recover the famous pseudo-Dirac scheme with maximal mixing. That is the eigenstates
will be the following Majorana states  with energy eigenvalues as
\begin{equation}
 \chi_1=\frac{\nu_L+\nu^c_R}{2} ~{\rm with}~ E^2=p^2+m^2 +\mu m
\end{equation}
and
\begin{equation}
\chi_2=\frac{\nu_L-\nu^c_R}{2} ~{\rm with}~ E^2=p^2+m^2 -\mu m
\end{equation}
As a result, active  $\nu_L$ can oscillate to sterile neutrino with oscillation length determined by splitting $\Delta m^2=2m\mu$ and maximal mixing
$$P(\nu_L\to \nu_R^c)=\sin^2(\frac{\mu m}{2p}L)$$
and
$$ P(\nu_L\to \nu_L)=1-P(\nu_L\to \nu_R^c)=\cos^2(\frac{\mu m}{2p}L)$$
Notice that the active and sterile neutrinos respectively correspond to $\nu_L$ and $\nu_R$ (or $\nu_R^c$). We therefore use $\nu_R^c$ and $\nu_s$ interchangeably: $P(\nu_L\to \nu_R^c)=P(\nu_a\to \nu_s)$.

For $2pV\gg \mu m$, the mixing between $\nu_L$ and $\nu_R^c$ will be suppressed by $\mu m/(2pV)$ so the oscillation to sterile neutrino will be negligible. For the sake of simplicity, for  the  three  neutrino flavors, we strict ourselves to the case that the $m$ and $\mu$ matrices can simultaneously be diagonalized.
Thus, in the mass basis all terms will be diagonal except for the effective potential $V\sum_{i,j}U_{ei}U_{ej}^* \bar{\Psi}_j \gamma^0 \Psi_i$. 

Now, let us consider solar neutrinos with $\mu m/(2p)\sim 1/L$ where $L$ is the Earth Sun distance. Within the Sun, the matter effects will dominate and will suppress the $\nu_{Li}$ and $\nu^c_{Ri}$ mixing. That is within the Sun, we shall have the standard MSW effect and the $\nu_e$ state after crossing the Sun will emerge at the Sun surface as an incoherent combination of $\nu_{L 1}$, $\nu_{L 2}$ and $\nu_{L 3}$ with pobabilities $\cos^2\theta_M\cos^2 \theta_{13}$, $\sin^2\theta_M\cos^2 \theta_{13}$ and $\sin^2 \theta_{13}$.
Here, $\theta_M$ is the effective 12 mixing at the production point of $\nu_e$:
\begin{align}\label{thetM}
\cos 2\theta_M & = \\ \nonumber
&\frac{\Delta m_{21}^2 \cos 2\theta_{12}-VE_\nu}{\left[ (\Delta m_{21}^2 \cos 2\theta_{12}-VE_\nu)^2+(\Delta m_{21}^2 \sin 2\theta_{12})^2\right]^{1/2}} 
\end{align}
in which $V=2\sqrt{2} G_F N_e(r)|_{\rm at ~production}$.
Notice that we have taken into account two facts: (i) conversion in the Sun is adiabatic; (ii) the matter effect on $\theta_{13}$ is negligible due to suppression  with  $2Vp/(\Delta m_{31}^2 \sin\theta_{13})\ll 1.$
The mass eigenstates on their way to Earth can oscillate into their sterile counter-part ($\nu^c_{Ri}$) with maximal mixing so the survival probability up to the Earth surface can be written as 
\begin{align}\label{Pee}
P_{ee}  = & \cos^4 \theta_{13}  \Biggl( \cos^2\theta_{12} \cos^2\theta_M \cos^2(\frac{\mu_1 m_1}{2p}L)  \\ \nonumber
& \qquad \qquad + \sin^2\theta_{12}  \sin^2\theta_M \cos^2(\frac{\mu_2 m_2}{2p}L) \Biggl) \\ \nonumber
& + \sin^4\theta_{13} \cos^2(\frac{\mu_3 m_3}{2p}L) 
\end{align}
and 
\begin{align}\label{Pes}
P_{es}  = &  \cos^2\theta_{13} \Biggl( \cos^2\theta_M\sin^2(\frac{\mu_1 m_1}{2p}L) \\ \nonumber  
& \qquad \qquad + \sin^2\theta_M\sin^2(\frac{\mu_2 m_2}{2p}L) \Biggl) \\ \nonumber
& +\sin^2 \theta_{13}\sin^2(\frac{\mu_3 m_3}{2p}L)
\end{align}  
For simplicity, we denote $P_{ee}\equiv P(\nu_e \to \nu_e)$ and $P_{es}\equiv \sum_i P(\nu_e\to \nu_{s_i})$.
This formula corresponds to that in Ref. \cite{deGouvea:2009fp} in the limit $\mu_im_i\ll VE_\nu$.
For relatively high energy solar neutrinos, the  oscillation in Earth due to matter effects ({\it i.e.,} Day/Night effect) can  also be important but our focus is on the intermediate energy solar neutrinos for which the matter effects are negligible.

Through $\theta_M$, $P_{ee}$ depends on the location of $\nu_e$ production inside the Sun.    We  define
\be \bar{P}^j_{ee}(E_\nu,L) = \int_0^{R_\odot} P_{ee}(E_\nu,L,r) \Phi_j(r)dr \ee
where $j$ can be any of the flux components $pp$, $^7$Be, $pep$ and $^8$B.  $\Phi_j(r)$ is the flux from radius $r$ taken from \footnote{http://www.sns.ias.edu/~jnb/}.
Notice that  $\Phi_j(r) dr$  includes the volume factor ($r^2 dr$) in its definition and vanishes at $r=0$. The dependence of $\Phi_j(r)$ on $r$ is different for the $j$ modes. For example, while for $j=pp$ the flux peaks at $r\simeq 0.1R_\odot$, for $j=^8$B, the peak is at  $r\simeq 0.05 R_\odot$, The dependence of $P_{ee}(E_\nu,L,r)$ on $r$ is through the dependence of $\theta_M$ on $N_e(r)$.
Let us define  
$$\Delta m_i^2=2\mu_i m_i . $$

Let us now discuss the time dependence of the flux throughout a year.
The sun Earth distance during a year varies between $L_{max}=152.1\times 10^6$~km (aphelion ocurring around July 4th) and $L_{min}=147.1\times 10^6$~km (perihelion ocurring around January 4th).  That is the orbit of the Earth around the Sun can be written as 
\be L(\theta)=\frac{a(1-e^2)}{1+e\cos\theta} \label{rtheta}\ee
in which $a=(L_{min}+L_{max})/2$ and the eccentricity is $e= (L_{max}-L_{min})/(L_{max}+L_{min})=0.0167$. The conservation of the angular momentum imples $dt =(L^2/H)d\theta$ in which $H=|\vec{r}\times \dot{\vec{r}}|$.
The  number of events during a time interval $(T_1,T_2)$ is proportional to 
\begin{align}
&\int_{t_1}^{t_1+\Delta t}\frac{P_{ee}\sigma_e+(1-P_{ee}-P_{es})\sigma_\mu}{L^2} dt = \\ \nonumber
& \int_{\theta_1}^{\theta_2}\frac{P_{ee}\sigma_e+(1-P_{ee}-P_{es})\sigma_\mu}{H} d\theta
\end{align}
where $\sigma_e$ and $\sigma_\mu$ are respectively the scattering cross sections of $\nu_e$ and $\nu_\mu$ (or $\nu_\tau$) at the detector.
To compute the number of events during a time interval we should know the relation between $\theta$ and time. Replacing $L(\theta)$ given in Eq. (\ref{rtheta}) we find
\begin{equation}
\int_{t_1}^{t_1+\Delta t} \frac{H}{a^2(1-e^2)^2}dt = \int_{\theta_1}^{\theta_2}\frac{d\theta}{(1+e\cos\theta)^2}
\end{equation}
which yields
$$H=\frac {6.55a^2(1-e^2)^2}{1~{\rm year}}.$$
We have used these formulas to study the seasonal variation of the $^7$Be solar flux.
As  shown in \cite{Bahcall:1994cf}, the widths of $^7$Be and $pep$ lines are of order of kinetic energy in the sun center $\Delta E_\nu\sim 0.6$ keV. Thus, as long as $\Delta m^2L/E\ll 1000$, we have $(\Delta E_\nu/E_\nu)(\Delta m^2L/E)\ll 1$, the  finite width of these lines will not smear
the oscillatory behavior.

\begin{acknowledgments}
The authors are  grateful to P. Zakeri for the collaboration in the early stages of this work. This project has received funding/support from the European Union’s Horizon 2020 research and innovation programme under the Marie Skłodowska -Curie grant agreement No 860881-HIDDeN. YF has received  financial support from Saramadan under contract No.~ISEF/M/401439. She would like to acknowledge support from the ICTP through the Associates Programme and from the Simons Foundation through grant number 284558FY19.
\end{acknowledgments}

\bibliography{PRD.bib}

\begin{thebibliography}{31}%
\makeatletter
\providecommand \@ifxundefined [1]{%
 \@ifx{#1\undefined}
}%
\providecommand \@ifnum [1]{%
 \ifnum #1\expandafter \@firstoftwo
 \else \expandafter \@secondoftwo
 \fi
}%
\providecommand \@ifx [1]{%
 \ifx #1\expandafter \@firstoftwo
 \else \expandafter \@secondoftwo
 \fi
}%
\providecommand \natexlab [1]{#1}%
\providecommand \enquote  [1]{``#1''}%
\providecommand \bibnamefont  [1]{#1}%
\providecommand \bibfnamefont [1]{#1}%
\providecommand \citenamefont [1]{#1}%
\providecommand \href@noop [0]{\@secondoftwo}%
\providecommand \href [0]{\begingroup \@sanitize@url \@href}%
\providecommand \@href[1]{\@@startlink{#1}\@@href}%
\providecommand \@@href[1]{\endgroup#1\@@endlink}%
\providecommand \@sanitize@url [0]{\catcode `\\12\catcode `\$12\catcode
  `\&12\catcode `\#12\catcode `\^12\catcode `\_12\catcode `\%12\relax}%
\providecommand \@@startlink[1]{}%
\providecommand \@@endlink[0]{}%
\providecommand \url  [0]{\begingroup\@sanitize@url \@url }%
\providecommand \@url [1]{\endgroup\@href {#1}{\urlprefix }}%
\providecommand \urlprefix  [0]{URL }%
\providecommand \Eprint [0]{\href }%
\providecommand \doibase [0]{https://doi.org/}%
\providecommand \selectlanguage [0]{\@gobble}%
\providecommand \bibinfo  [0]{\@secondoftwo}%
\providecommand \bibfield  [0]{\@secondoftwo}%
\providecommand \translation [1]{[#1]}%
\providecommand \BibitemOpen [0]{}%
\providecommand \bibitemStop [0]{}%
\providecommand \bibitemNoStop [0]{.\EOS\space}%
\providecommand \EOS [0]{\spacefactor3000\relax}%
\providecommand \BibitemShut  [1]{\csname bibitem#1\endcsname}%
\let\auto@bib@innerbib\@empty
\bibitem [{\citenamefont {Anamiati}\ \emph {et~al.}(2019)\citenamefont
  {Anamiati}, \citenamefont {De~Romeri}, \citenamefont {Hirsch}, \citenamefont
  {Ternes},\ and\ \citenamefont {T\'ortola}}]{Anamiati:2019maf}%
  \BibitemOpen
  \bibfield  {author} {\bibinfo {author} {\bibfnamefont {G.}~\bibnamefont
  {Anamiati}}, \bibinfo {author} {\bibfnamefont {V.}~\bibnamefont {De~Romeri}},
  \bibinfo {author} {\bibfnamefont {M.}~\bibnamefont {Hirsch}}, \bibinfo
  {author} {\bibfnamefont {C.~A.}\ \bibnamefont {Ternes}},\ and\ \bibinfo
  {author} {\bibfnamefont {M.}~\bibnamefont {T\'ortola}},\ }\href
  {https://doi.org/10.1103/PhysRevD.100.035032} {\bibfield  {journal} {\bibinfo
   {journal} {Phys. Rev. D}\ }\textbf {\bibinfo {volume} {100}},\ \bibinfo
  {pages} {035032} (\bibinfo {year} {2019})},\ \Eprint
  {https://arxiv.org/abs/1907.00980} {arXiv:1907.00980 [hep-ph]} \BibitemShut
  {NoStop}%
\bibitem [{\citenamefont {Martinez-Soler}\ \emph {et~al.}(2022)\citenamefont
  {Martinez-Soler}, \citenamefont {Perez-Gonzalez},\ and\ \citenamefont
  {Sen}}]{Martinez-Soler:2021unz}%
  \BibitemOpen
  \bibfield  {author} {\bibinfo {author} {\bibfnamefont {I.}~\bibnamefont
  {Martinez-Soler}}, \bibinfo {author} {\bibfnamefont {Y.~F.}\ \bibnamefont
  {Perez-Gonzalez}},\ and\ \bibinfo {author} {\bibfnamefont {M.}~\bibnamefont
  {Sen}},\ }\href {https://doi.org/10.1103/PhysRevD.105.095019} {\bibfield
  {journal} {\bibinfo  {journal} {Phys. Rev. D}\ }\textbf {\bibinfo {volume}
  {105}},\ \bibinfo {pages} {095019} (\bibinfo {year} {2022})},\ \Eprint
  {https://arxiv.org/abs/2105.12736} {arXiv:2105.12736 [hep-ph]} \BibitemShut
  {NoStop}%
\bibitem [{\citenamefont {Esmaili}\ and\ \citenamefont
  {Farzan}(2012)}]{Esmaili:2012ac}%
  \BibitemOpen
  \bibfield  {author} {\bibinfo {author} {\bibfnamefont {A.}~\bibnamefont
  {Esmaili}}\ and\ \bibinfo {author} {\bibfnamefont {Y.}~\bibnamefont
  {Farzan}},\ }\href {https://doi.org/10.1088/1475-7516/2012/12/014} {\bibfield
   {journal} {\bibinfo  {journal} {JCAP}\ }\textbf {\bibinfo {volume} {12}},\
  \bibinfo {pages} {014}},\ \Eprint {https://arxiv.org/abs/1208.6012}
  {arXiv:1208.6012 [hep-ph]} \BibitemShut {NoStop}%
\bibitem [{\citenamefont {Joshipura}\ \emph {et~al.}(2014)\citenamefont
  {Joshipura}, \citenamefont {Mohanty},\ and\ \citenamefont
  {Pakvasa}}]{Joshipura:2013yba}%
  \BibitemOpen
  \bibfield  {author} {\bibinfo {author} {\bibfnamefont {A.~S.}\ \bibnamefont
  {Joshipura}}, \bibinfo {author} {\bibfnamefont {S.}~\bibnamefont {Mohanty}},\
  and\ \bibinfo {author} {\bibfnamefont {S.}~\bibnamefont {Pakvasa}},\ }\href
  {https://doi.org/10.1103/PhysRevD.89.033003} {\bibfield  {journal} {\bibinfo
  {journal} {Phys. Rev. D}\ }\textbf {\bibinfo {volume} {89}},\ \bibinfo
  {pages} {033003} (\bibinfo {year} {2014})},\ \Eprint
  {https://arxiv.org/abs/1307.5712} {arXiv:1307.5712 [hep-ph]} \BibitemShut
  {NoStop}%
\bibitem [{\citenamefont {Crocker}\ \emph {et~al.}(2000)\citenamefont
  {Crocker}, \citenamefont {Melia},\ and\ \citenamefont
  {Volkas}}]{Crocker:1999yw}%
  \BibitemOpen
  \bibfield  {author} {\bibinfo {author} {\bibfnamefont {R.~M.}\ \bibnamefont
  {Crocker}}, \bibinfo {author} {\bibfnamefont {F.}~\bibnamefont {Melia}},\
  and\ \bibinfo {author} {\bibfnamefont {R.~R.}\ \bibnamefont {Volkas}},\
  }\href {https://doi.org/10.1086/317350} {\bibfield  {journal} {\bibinfo
  {journal} {Astrophys. J. Suppl.}\ }\textbf {\bibinfo {volume} {130}},\
  \bibinfo {pages} {339} (\bibinfo {year} {2000})},\ \Eprint
  {https://arxiv.org/abs/astro-ph/9911292} {arXiv:astro-ph/9911292}
  \BibitemShut {NoStop}%
\bibitem [{\citenamefont {Esmaili}(2010)}]{Esmaili:2009fk}%
  \BibitemOpen
  \bibfield  {author} {\bibinfo {author} {\bibfnamefont {A.}~\bibnamefont
  {Esmaili}},\ }\href {https://doi.org/10.1103/PhysRevD.81.013006} {\bibfield
  {journal} {\bibinfo  {journal} {Phys. Rev. D}\ }\textbf {\bibinfo {volume}
  {81}},\ \bibinfo {pages} {013006} (\bibinfo {year} {2010})},\ \Eprint
  {https://arxiv.org/abs/0909.5410} {arXiv:0909.5410 [hep-ph]} \BibitemShut
  {NoStop}%
\bibitem [{\citenamefont {Keranen}\ \emph {et~al.}(2003)\citenamefont
  {Keranen}, \citenamefont {Maalampi}, \citenamefont {Myyrylainen},\ and\
  \citenamefont {Riittinen}}]{Keranen:2003xd}%
  \BibitemOpen
  \bibfield  {author} {\bibinfo {author} {\bibfnamefont {P.}~\bibnamefont
  {Keranen}}, \bibinfo {author} {\bibfnamefont {J.}~\bibnamefont {Maalampi}},
  \bibinfo {author} {\bibfnamefont {M.}~\bibnamefont {Myyrylainen}},\ and\
  \bibinfo {author} {\bibfnamefont {J.}~\bibnamefont {Riittinen}},\ }\href
  {https://doi.org/10.1016/j.physletb.2003.09.006} {\bibfield  {journal}
  {\bibinfo  {journal} {Phys. Lett. B}\ }\textbf {\bibinfo {volume} {574}},\
  \bibinfo {pages} {162} (\bibinfo {year} {2003})},\ \Eprint
  {https://arxiv.org/abs/hep-ph/0307041} {arXiv:hep-ph/0307041} \BibitemShut
  {NoStop}%
\bibitem [{\citenamefont {Crocker}\ \emph {et~al.}(2002)\citenamefont
  {Crocker}, \citenamefont {Melia},\ and\ \citenamefont
  {Volkas}}]{Crocker:2001zs}%
  \BibitemOpen
  \bibfield  {author} {\bibinfo {author} {\bibfnamefont {R.~M.}\ \bibnamefont
  {Crocker}}, \bibinfo {author} {\bibfnamefont {F.}~\bibnamefont {Melia}},\
  and\ \bibinfo {author} {\bibfnamefont {R.~R.}\ \bibnamefont {Volkas}},\
  }\href {https://doi.org/10.1086/340278} {\bibfield  {journal} {\bibinfo
  {journal} {Astrophys. J. Suppl.}\ }\textbf {\bibinfo {volume} {141}},\
  \bibinfo {pages} {147} (\bibinfo {year} {2002})},\ \Eprint
  {https://arxiv.org/abs/astro-ph/0106090} {arXiv:astro-ph/0106090}
  \BibitemShut {NoStop}%
\bibitem [{\citenamefont {Anamiati}\ \emph {et~al.}(2018)\citenamefont
  {Anamiati}, \citenamefont {Fonseca},\ and\ \citenamefont
  {Hirsch}}]{Anamiati:2017rxw}%
  \BibitemOpen
  \bibfield  {author} {\bibinfo {author} {\bibfnamefont {G.}~\bibnamefont
  {Anamiati}}, \bibinfo {author} {\bibfnamefont {R.~M.}\ \bibnamefont
  {Fonseca}},\ and\ \bibinfo {author} {\bibfnamefont {M.}~\bibnamefont
  {Hirsch}},\ }\href {https://doi.org/10.1103/PhysRevD.97.095008} {\bibfield
  {journal} {\bibinfo  {journal} {Phys. Rev. D}\ }\textbf {\bibinfo {volume}
  {97}},\ \bibinfo {pages} {095008} (\bibinfo {year} {2018})},\ \Eprint
  {https://arxiv.org/abs/1710.06249} {arXiv:1710.06249 [hep-ph]} \BibitemShut
  {NoStop}%
\bibitem [{\citenamefont {de~Gouvea}\ \emph {et~al.}(2009)\citenamefont
  {de~Gouvea}, \citenamefont {Huang},\ and\ \citenamefont
  {Jenkins}}]{deGouvea:2009fp}%
  \BibitemOpen
  \bibfield  {author} {\bibinfo {author} {\bibfnamefont {A.}~\bibnamefont
  {de~Gouvea}}, \bibinfo {author} {\bibfnamefont {W.-C.}\ \bibnamefont
  {Huang}},\ and\ \bibinfo {author} {\bibfnamefont {J.}~\bibnamefont
  {Jenkins}},\ }\href {https://doi.org/10.1103/PhysRevD.80.073007} {\bibfield
  {journal} {\bibinfo  {journal} {Phys. Rev. D}\ }\textbf {\bibinfo {volume}
  {80}},\ \bibinfo {pages} {073007} (\bibinfo {year} {2009})},\ \Eprint
  {https://arxiv.org/abs/0906.1611} {arXiv:0906.1611 [hep-ph]} \BibitemShut
  {NoStop}%
\bibitem [{\citenamefont {Esteban}\ \emph {et~al.}(2020)\citenamefont
  {Esteban}, \citenamefont {Gonzalez-Garcia}, \citenamefont {Maltoni},
  \citenamefont {Schwetz},\ and\ \citenamefont {Zhou}}]{Esteban:2020cvm}%
  \BibitemOpen
  \bibfield  {author} {\bibinfo {author} {\bibfnamefont {I.}~\bibnamefont
  {Esteban}}, \bibinfo {author} {\bibfnamefont {M.~C.}\ \bibnamefont
  {Gonzalez-Garcia}}, \bibinfo {author} {\bibfnamefont {M.}~\bibnamefont
  {Maltoni}}, \bibinfo {author} {\bibfnamefont {T.}~\bibnamefont {Schwetz}},\
  and\ \bibinfo {author} {\bibfnamefont {A.}~\bibnamefont {Zhou}},\ }\href
  {https://doi.org/10.1007/JHEP09(2020)178} {\bibfield  {journal} {\bibinfo
  {journal} {JHEP}\ }\textbf {\bibinfo {volume} {09}},\ \bibinfo {pages}
  {178}},\ \Eprint {https://arxiv.org/abs/2007.14792} {arXiv:2007.14792
  [hep-ph]} \BibitemShut {NoStop}%
\bibitem [{\citenamefont {Gando}\ \emph {et~al.}(2011)\citenamefont {Gando}
  \emph {et~al.}}]{KamLAND:2010fvi}%
  \BibitemOpen
  \bibfield  {author} {\bibinfo {author} {\bibfnamefont {A.}~\bibnamefont
  {Gando}} \emph {et~al.} (\bibinfo {collaboration} {KamLAND}),\ }\href
  {https://doi.org/10.1103/PhysRevD.83.052002} {\bibfield  {journal} {\bibinfo
  {journal} {Phys. Rev. D}\ }\textbf {\bibinfo {volume} {83}},\ \bibinfo
  {pages} {052002} (\bibinfo {year} {2011})},\ \Eprint
  {https://arxiv.org/abs/1009.4771} {arXiv:1009.4771 [hep-ex]} \BibitemShut
  {NoStop}%
\bibitem [{\citenamefont {Abe}\ \emph {et~al.}(2016)\citenamefont {Abe} \emph
  {et~al.}}]{Super-Kamiokande:2016yck}%
  \BibitemOpen
  \bibfield  {author} {\bibinfo {author} {\bibfnamefont {K.}~\bibnamefont
  {Abe}} \emph {et~al.} (\bibinfo {collaboration} {Super-Kamiokande}),\ }\href
  {https://doi.org/10.1103/PhysRevD.94.052010} {\bibfield  {journal} {\bibinfo
  {journal} {Phys. Rev. D}\ }\textbf {\bibinfo {volume} {94}},\ \bibinfo
  {pages} {052010} (\bibinfo {year} {2016})},\ \Eprint
  {https://arxiv.org/abs/1606.07538} {arXiv:1606.07538 [hep-ex]} \BibitemShut
  {NoStop}%
\bibitem [{Note1()}]{Note1}%
  \BibitemOpen
  \bibinfo {note} {Regarding to day/night effect see the Appendix.}\BibitemShut
  {Stop}%
\bibitem [{\citenamefont {Nakajima}(2020)}]{Super-Kamiokande:2020}%
  \BibitemOpen
  \bibfield  {author} {\bibinfo {author} {\bibfnamefont {Y.}~\bibnamefont
  {Nakajima}} (\bibinfo {collaboration} {Super-Kamiokande}),\ }\href@noop {}
  {\bibfield  {journal} {\bibinfo  {journal} {Talk given at the XXIX
  International Conference on Neutrino Physics and Astrophysics}\ ,\ \bibinfo
  {pages} {June 30}} (\bibinfo {year} {2020})}\BibitemShut {NoStop}%
\bibitem [{\citenamefont {Agostini}\ \emph {et~al.}(2020)\citenamefont
  {Agostini} \emph {et~al.}}]{BOREXINO:2020hox}%
  \BibitemOpen
  \bibfield  {author} {\bibinfo {author} {\bibfnamefont {M.}~\bibnamefont
  {Agostini}} \emph {et~al.} (\bibinfo {collaboration} {BOREXINO}),\ }\href
  {https://doi.org/10.1140/epjc/s10052-020-08534-2} {\bibfield  {journal}
  {\bibinfo  {journal} {Eur. Phys. J. C}\ }\textbf {\bibinfo {volume} {80}},\
  \bibinfo {pages} {1091} (\bibinfo {year} {2020})},\ \Eprint
  {https://arxiv.org/abs/2005.12829} {arXiv:2005.12829 [hep-ex]} \BibitemShut
  {NoStop}%
\bibitem [{\citenamefont {Vinyoles}\ \emph {et~al.}(2017)\citenamefont
  {Vinyoles}, \citenamefont {Serenelli}, \citenamefont {Villante},
  \citenamefont {Basu}, \citenamefont {Bergstr\"om}, \citenamefont
  {Gonzalez-Garcia}, \citenamefont {Maltoni}, \citenamefont {Pe\~na Garay},\
  and\ \citenamefont {Song}}]{Vinyoles:2016djt}%
  \BibitemOpen
  \bibfield  {author} {\bibinfo {author} {\bibfnamefont {N.}~\bibnamefont
  {Vinyoles}}, \bibinfo {author} {\bibfnamefont {A.~M.}\ \bibnamefont
  {Serenelli}}, \bibinfo {author} {\bibfnamefont {F.~L.}\ \bibnamefont
  {Villante}}, \bibinfo {author} {\bibfnamefont {S.}~\bibnamefont {Basu}},
  \bibinfo {author} {\bibfnamefont {J.}~\bibnamefont {Bergstr\"om}}, \bibinfo
  {author} {\bibfnamefont {M.~C.}\ \bibnamefont {Gonzalez-Garcia}}, \bibinfo
  {author} {\bibfnamefont {M.}~\bibnamefont {Maltoni}}, \bibinfo {author}
  {\bibfnamefont {C.}~\bibnamefont {Pe\~na Garay}},\ and\ \bibinfo {author}
  {\bibfnamefont {N.}~\bibnamefont {Song}},\ }\href
  {https://doi.org/10.3847/1538-4357/835/2/202} {\bibfield  {journal} {\bibinfo
   {journal} {Astrophys. J.}\ }\textbf {\bibinfo {volume} {835}},\ \bibinfo
  {pages} {202} (\bibinfo {year} {2017})},\ \Eprint
  {https://arxiv.org/abs/1611.09867} {arXiv:1611.09867 [astro-ph.SR]}
  \BibitemShut {NoStop}%
\bibitem [{\citenamefont {Nakano}(2016)}]{Super-PhD}%
  \BibitemOpen
  \bibfield  {author} {\bibinfo {author} {\bibfnamefont {Y.}~\bibnamefont
  {Nakano}} (\bibinfo {collaboration} {Super-Kamiokande}),\ }\href@noop {}
  {\bibfield  {journal} {\bibinfo  {journal} {PhD thesis}\ ,\ \bibinfo {pages}
  {Tokyo U.}} (\bibinfo {year} {2016})}\BibitemShut {NoStop}%
\bibitem [{\citenamefont {Chen}\ \emph {et~al.}(2021)\citenamefont {Chen},
  \citenamefont {Li},\ and\ \citenamefont {Liao}}]{Chen:2021uuw}%
  \BibitemOpen
  \bibfield  {author} {\bibinfo {author} {\bibfnamefont {Z.}~\bibnamefont
  {Chen}}, \bibinfo {author} {\bibfnamefont {T.}~\bibnamefont {Li}},\ and\
  \bibinfo {author} {\bibfnamefont {J.}~\bibnamefont {Liao}},\ }\href
  {https://doi.org/10.1007/JHEP05(2021)131} {\bibfield  {journal} {\bibinfo
  {journal} {JHEP}\ }\textbf {\bibinfo {volume} {05}},\ \bibinfo {pages}
  {131}},\ \Eprint {https://arxiv.org/abs/2102.09784} {arXiv:2102.09784
  [hep-ph]} \BibitemShut {NoStop}%
\bibitem [{Note2()}]{Note2}%
  \BibitemOpen
  \bibinfo {note} {\protect \url {https://pdg.lbl.gov/}}\BibitemShut {NoStop}%
\bibitem [{\citenamefont {Altmann}\ \emph {et~al.}(2005)\citenamefont {Altmann}
  \emph {et~al.}}]{GNO:2005bds}%
  \BibitemOpen
  \bibfield  {author} {\bibinfo {author} {\bibfnamefont {M.}~\bibnamefont
  {Altmann}} \emph {et~al.} (\bibinfo {collaboration} {GNO}),\ }\href
  {https://doi.org/10.1016/j.physletb.2005.04.068} {\bibfield  {journal}
  {\bibinfo  {journal} {Phys. Lett. B}\ }\textbf {\bibinfo {volume} {616}},\
  \bibinfo {pages} {174} (\bibinfo {year} {2005})},\ \Eprint
  {https://arxiv.org/abs/hep-ex/0504037} {arXiv:hep-ex/0504037} \BibitemShut
  {NoStop}%
\bibitem [{Note3()}]{Note3}%
  \BibitemOpen
  \bibinfo {note} {We have also performed a similar analysis with the official
  Super-Kamiokande data release in 2016 \cite {Super-Kamiokande:2016yck} which
  confirmed the current results, except that those data tended to have lower
  values and thus the region found in the interval of $\Delta m_{2}^2 =
  (1,2)\times 10^{-11}$ compared to the $\Delta m_{2}^2 \to 0$ have lower $\chi
  ^2$ value; {\protect \it i.e.,} providing a slightly better fit than the
  standard MSW with $\Delta m_{2}^2 =0$.}\BibitemShut {Stop}%
\bibitem [{\citenamefont {Aharmim}\ \emph
  {et~al.}(2013{\natexlab{a}})\citenamefont {Aharmim} \emph
  {et~al.}}]{SNO:2011hxd}%
  \BibitemOpen
  \bibfield  {author} {\bibinfo {author} {\bibfnamefont {B.}~\bibnamefont
  {Aharmim}} \emph {et~al.} (\bibinfo {collaboration} {SNO}),\ }\href
  {https://doi.org/10.1103/PhysRevC.88.025501} {\bibfield  {journal} {\bibinfo
  {journal} {Phys. Rev. C}\ }\textbf {\bibinfo {volume} {88}},\ \bibinfo
  {pages} {025501} (\bibinfo {year} {2013}{\natexlab{a}})},\ \Eprint
  {https://arxiv.org/abs/1109.0763} {arXiv:1109.0763 [nucl-ex]} \BibitemShut
  {NoStop}%
\bibitem [{\citenamefont {Aharmim}\ \emph
  {et~al.}(2013{\natexlab{b}})\citenamefont {Aharmim} \emph
  {et~al.}}]{SNO:2011ajh}%
  \BibitemOpen
  \bibfield  {author} {\bibinfo {author} {\bibfnamefont {B.}~\bibnamefont
  {Aharmim}} \emph {et~al.} (\bibinfo {collaboration} {SNO}),\ }\href
  {https://doi.org/10.1103/PhysRevC.87.015502} {\bibfield  {journal} {\bibinfo
  {journal} {Phys. Rev. C}\ }\textbf {\bibinfo {volume} {87}},\ \bibinfo
  {pages} {015502} (\bibinfo {year} {2013}{\natexlab{b}})},\ \Eprint
  {https://arxiv.org/abs/1107.2901} {arXiv:1107.2901 [nucl-ex]} \BibitemShut
  {NoStop}%
\bibitem [{Note4()}]{Note4}%
  \BibitemOpen
  \bibinfo {note} {The natural energy threshold, which is set by the binding
  energy of the Deuteron nucleus, is 2.2 MeV.}\BibitemShut {Stop}%
\bibitem [{\citenamefont {Appel}\ \emph {et~al.}(2023)\citenamefont {Appel}
  \emph {et~al.}}]{BOREXINO:2022wuy}%
  \BibitemOpen
  \bibfield  {author} {\bibinfo {author} {\bibfnamefont {S.}~\bibnamefont
  {Appel}} \emph {et~al.} (\bibinfo {collaboration} {BOREXINO}),\ }\href
  {https://doi.org/10.1016/j.astropartphys.2022.102778} {\bibfield  {journal}
  {\bibinfo  {journal} {Astropart. Phys.}\ }\textbf {\bibinfo {volume} {145}},\
  \bibinfo {pages} {102778} (\bibinfo {year} {2023})},\ \Eprint
  {https://arxiv.org/abs/2204.07029} {arXiv:2204.07029 [hep-ex]} \BibitemShut
  {NoStop}%
\bibitem [{Note5()}]{Note5}%
  \BibitemOpen
  \bibinfo {note} {For the exact definition of tend and residue, the readers
  may consult \cite {BOREXINO:2022wuy}.}\BibitemShut {Stop}%
\bibitem [{\citenamefont {Tegmark}\ \emph {et~al.}(1997)\citenamefont
  {Tegmark}, \citenamefont {Taylor},\ and\ \citenamefont
  {Heavens}}]{Tegmark:1996bz}%
  \BibitemOpen
  \bibfield  {author} {\bibinfo {author} {\bibfnamefont {M.}~\bibnamefont
  {Tegmark}}, \bibinfo {author} {\bibfnamefont {A.}~\bibnamefont {Taylor}},\
  and\ \bibinfo {author} {\bibfnamefont {A.}~\bibnamefont {Heavens}},\ }\href
  {https://doi.org/10.1086/303939} {\bibfield  {journal} {\bibinfo  {journal}
  {Astrophys. J.}\ }\textbf {\bibinfo {volume} {480}},\ \bibinfo {pages} {22}
  (\bibinfo {year} {1997})},\ \Eprint {https://arxiv.org/abs/astro-ph/9603021}
  {arXiv:astro-ph/9603021} \BibitemShut {NoStop}%
\bibitem [{\citenamefont {Askins}\ \emph {et~al.}(2020)\citenamefont {Askins}
  \emph {et~al.}}]{Theia:2019non}%
  \BibitemOpen
  \bibfield  {author} {\bibinfo {author} {\bibfnamefont {M.}~\bibnamefont
  {Askins}} \emph {et~al.} (\bibinfo {collaboration} {Theia}),\ }\href
  {https://doi.org/10.1140/epjc/s10052-020-7977-8} {\bibfield  {journal}
  {\bibinfo  {journal} {Eur. Phys. J. C}\ }\textbf {\bibinfo {volume} {80}},\
  \bibinfo {pages} {416} (\bibinfo {year} {2020})},\ \Eprint
  {https://arxiv.org/abs/1911.03501} {arXiv:1911.03501 [physics.ins-det]}
  \BibitemShut {NoStop}%
\bibitem [{Note6()}]{Note6}%
  \BibitemOpen
  \bibinfo {note} {Http://www.sns.ias.edu/~jnb/}\BibitemShut {NoStop}%
\bibitem [{\citenamefont {Bahcall}(1994)}]{Bahcall:1994cf}%
  \BibitemOpen
  \bibfield  {author} {\bibinfo {author} {\bibfnamefont {J.~N.}\ \bibnamefont
  {Bahcall}},\ }\href {https://doi.org/10.1103/PhysRevD.49.3923} {\bibfield
  {journal} {\bibinfo  {journal} {Phys. Rev. D}\ }\textbf {\bibinfo {volume}
  {49}},\ \bibinfo {pages} {3923} (\bibinfo {year} {1994})},\ \Eprint
  {https://arxiv.org/abs/astro-ph/9401024} {arXiv:astro-ph/9401024}
  \BibitemShut {NoStop}%
\end{thebibliography}%

\end{document}